\documentclass[a4paper,fleqn,usenatbib]{mn2e}

\usepackage{aas_macros}
\usepackage{color}
\usepackage{bm}
\usepackage{hyperref}
\usepackage{newtxtext,newtxmath}
\usepackage[T1]{fontenc}
\usepackage{times}
\usepackage{amssymb}
\usepackage{float}
\usepackage{graphicx}
\usepackage{epstopdf}
\usepackage{lscape}
\usepackage{threeparttable}
\usepackage{bigints}

\def\kms{km s$^{-1}$ }
\newcommand{\Gaia}{{\it Gaia }}

\title[Predicting the HVS population in \Gaia]{Predicting the hypervelocity star population in \Gaia}

\author[T. Marchetti et al.]{T. Marchetti$^{1}$\thanks{E-mail: marchetti@strw.leidenuniv.nl},  O. Contigiani$^1$, E. M. Rossi$^{1}$, J. G. Albert$^{1}$, A. G. A. Brown$^{1}$ \newauthor and A. Sesana$^2$ \\
$^1$Leiden Observatory, Leiden University, PO Box 9513 2300 RA Leiden, the Netherlands \\
$^2$School of Physics and Astronomy, University of Birmingham, Edgbaston, Birmingham 
B15 2TT, United Kingdom \\
}

\begin{document}

\pagerange{\pageref{firstpage}--\pageref{lastpage}} \pubyear{2017}

\maketitle

\label{firstpage}

\begin{abstract}

Hypervelocity stars (HVSs) are amongst the fastest objects in our Milky Way. These stars are predicted to come from the Galactic center (GC) and travel along unbound orbits across the Galaxy. In the coming years, the ESA satellite \Gaia will provide the most complete and accurate catalogue of the Milky Way, with full astrometric parameters for more than $1$ billion stars. In this paper, we present the expected sample size and properties (mass, magnitude, spatial, velocity distributions) of HVSs in the \Gaia stellar catalogue. We build three \Gaia mock catalogues of HVSs anchored to current observations, exploring different ejection mechanisms and GC stellar population properties. In all cases, we predict \emph{hundreds} to \emph{thousands} of HVSs with precise proper motion measurements within a few tens of kpc from us. For stars with a relative error in total proper motion below $10 \%$, the mass range extends to $\sim10 M_{\odot}$ but peaks at $\sim 1$ $M_\odot$. The majority of Gaia HVSs will therefore probe a different mass and distance range compared to the current non-\Gaia sample. In addition, a subset of a few hundreds to a few thousands of HVSs with $M \sim 3$ M$_\odot$ will be bright enough to have a precise measurement of the three-dimensional velocity from \Gaia alone.
Finally, we show that \Gaia will provide more precise proper motion measurements for the current sample of HVS candidates. This will help identifying their birthplace narrowing down their ejection location, and confirming or rejecting their nature as HVSs. Overall, our forecasts are extremely encouraging in terms of quantity and quality of HVS data that can be exploited to constrain both the Milky Way potential and the GC properties.

\end{abstract}


\section{Introduction}
\label{sec:intro}

A hypervelocity star (HVS) is a star observationally characterized by two main properties: its velocity is higher than the local escape velocity from our Galaxy (it is gravitationally unbound), and its orbit is consistent with a Galactocentric origin \citep{brown15}. The term HVS was originally coined by \cite{hills88}, and the first detection happened only in 2005 \citep{brown+05}. Currently $\sim 20$ HVS candidates have been found by the MMT HVS Survey of the northern hemisphere, in a mass range $[2.5, 4]$ M$_\odot$, and at distances between $50$ kpc and $100$ kpc from the Galactic Centre (GC) \citep{Brown+14}. This restricted mass range is an observational bias due to the survey detection strategy, that targets massive late B-type stars in the outer halo, that were not supposed to be found there (the halo is not a region of active star formation), unless they were ejected somewhere else with very high velocities. Lower mass HVSs have been searched for in the inner Galactic halo, using high proper motion, high radial velocity, and/or metallicity criteria. Most of these candidates are bound to the Galaxy, and/or their trajectories seem to be consistent with a Galactic disc origin (e.g. \citep{heber+08, palladino+14, zheng+14, hawkins+15, ziegerer+15, zhang+16, ziegerer+17}).

One puzzling aspect of the observed sample of B-type HVSs is their sky distribution: about half of the candidates are clumped in a small region of the sky ($5$ \% of the coverage area of the MMT HVS Survey), in the direction of the Leo constellation \citep{brown15}. Different ejection mechanisms predict different distributions of HVSs in the sky, and a full sky survey is needed in order to identify the physics responsible for their acceleration. 

The leading mechanism to explain the acceleration of a star up to $\sim 1000$ \kms  is the Hills mechanism \citep{hills88}. According to this scenario, HVSs are the result of a three body interaction between a binary star and the massive black hole (MBH) residing in the centre of our Galaxy, Sagittarius A*. In it simpler version, this mechanism predicts an isotropic distribution of HVSs in the sky. One possible alternative ejection mechanism involves the interaction of a single star with a massive black hole binary (MBHB) in the GC \citep{yu&tremaine03}.  Current observations cannot exclude the presence of a secondary massive compact object companion to Sagittarius A$^*$, with present upper limits around $10^4$ M$_\odot$ \citep{gillessen+17}. In this case, the ejection of HVSs becomes more energetic as the binary shrinks, and it  typical lasts for tens of Myr. This results in a ring of HVSs ejected in a very short burst, compared to the continuous ejection of stars predicted by the Hills mechanism (e.g. \citep{gualandris+05, sesana+06, sesana+08}). Other mechanisms involve the interaction of a globular cluster with a super massive black hole \citep{capuzzodolcetta+15} or with a MBHB \citep{fragione+16}, the interaction between a single star and a stellar black hole orbiting a MBH \citep{oleary+08}, and the tidal disruption of a dwarf galaxy \citep{abadi+09}. Recent observations have even shown evidence of star formation inside a galactic outflow ejected with high velocity from an active galactic nucleus \citep{maiolino+17}, suggesting that HVSs can be produced in other galaxies in such jets \citep{silk+12, zubovas+13}.

A more recent explanation for the observed B-type HVSs is given by \cite{boubert+17}, which interpret the current sample of candidates clumped in the direction of the Leo constellation as runaway stars from the Large Magellanic Cloud (LMC). 
Alternatively, HVSs could be produced by an hypothetical MBH in the centre of the LMC with a process that is analogous to the Hills mechanism \citep{boubert+16}.

All these mechanisms predict an additional population of stars, called \emph{bound HVSs}. These objects are formed in the same scenario as HVSs, but their velocity is not sufficiently high to escape from the gravitational field of the MW \citep[e.g.][]{bromley+06, kenyon+08}. These slower stars can travel along a wide variety of orbits, making their identification very difficult \citep{marchetti+17}.

In the past years HVSs have been proposed as tools to study multiple components of our Galaxy. The orbits of HVSs, spanning an unprecedented range of distances from the GC, integrate the Galactic potential, making them powerful tracers to study the matter distribution and orientation of the MW (i.e. \cite{gnedin+05, sesana+07, yu&madau07, kenyon+14, fragione&loeb16}). On the other hand, HVSs come from the GC, therefore they can be used to probe the stellar population near a quiescent MBH \citep{kollmeier+09, kollmeier+10}. It has been shown that a fraction of the original companions of HVSs can be tidally disrupted by the MBH, therefore the ejection rate of HVSs is directly linked to the growth rate of Sagittarius A$^*$ \citep{bromley+12}. A clean sample of HVSs would be also useful to constrain the metallicity distribution of stars in the GC. \cite{Rossi+17}, adopting the Hills mechanism, first attempted to constrain both the properties of the binary population in the GC (in terms of distributions of semi-major axes and mass ratios) and the scale parameters of the dark matter halo, using the sample of unbound HVSs from \cite{Brown+14}. They show that degeneracies between the parameters are preventing us from giving tight constraints, because of both the restricted number and the small mass range of the HVS candidates.

The ESA satellite \Gaia is going to revolutionize our knowledge of HVSs, shining a new light on their properties and origin. Launched in 2013, \Gaia is currently mapping the sky with an unprecedented accuracy, and by its final release (the end of 2022) it will provide precise positions, magnitudes, colours, parallaxes, and proper motions for more than $1$ billion stars \citep{gaiaa, gaiab}. Moreover, the Radial Velocity Spectrometer (RVS) on board will measure radial velocities for a subset of bright stars (magnitude in the \Gaia RVS band $G_\mathrm{RVS} < 16$). On the 14th September 2016 the first data  (\Gaia DR1) were released. The catalogue contains positions and $G$ magnitudes for more than 1 billion of sources. In addition, the five parameter astrometric solution (position, parallax, and proper motions) is available for a subset of $\sim 2 \times 10^6$ stars in common between \Gaia and the Tycho-2 catalogue: the Tycho-\Gaia Astrometric Solution (TGAS) catalogue \citep{michalik+15, lindegren+16}. The next data release, \Gaia DR2, is planned for the 25th of April 2018, and will be consisting of the five parameter astrometric solution, magnitudes, and colours for the full sample of stars ($> 10^9$ sources). It will also provide radial velocities for $5$ to $7$ million stars brighter than the $12$th magnitude in the $G_\mathrm{RVS}$ band. Effective temperatures, line-of-sight extinctions, luminosities, and radii will be provided for stars brighter than the $17$th magnitude in the $G$ band \citep{katz+17}.

A first attempt to find HVSs in \Gaia DR1/TGAS can be found in \cite{marchetti+17}, who developed a data-mining routine based on an artificial neural network trained on mock populations to distinguish HVSs from the dominant background of other stars in the Milky Way, using only the provided astrometry and no radial velocity information. This approach avoids biasing the search for HVSs towards particular spectral types, making as few assumptions as possible on the expected stellar properties. They found a total of $14$ stars with a total velocity in the Galactic rest frame higher than $400$ \kms, but because of large uncertainties, a clear identification of these candidates as HVSs is still uncertain. Five of these stars have a probability higher than $50$\% of being unbound from the MW. Because most of the stars have masses of the order of the Solar mass, they form a different population compared to the observed late B-type stars.

In this work, we forecast the sample size and properties of the HVS data expected in the next data releases of \Gaia, starting in April with DR2. The manuscript is organised as follows. In Section \ref{sec:mock} we explain how we build our first mock catalogue of HVSs, the \textsc{Vesc} catalogue, using a simple assumption on the total stellar velocity, and how we simulate \Gaia observations of these stars. Here we present our first results: how many HVSs we are expecting to find in the \Gaia catalogue using this first simple catalogue. In Section \ref{sec:Hills} we specialise our estimates on HVSs adopting the Hills mechanism, drawing velocities from a probability distribution, and we show how previous estimates and results change because of this assumption. In Section \ref{sec:MBHB} we build the third mock catalogue, the \textsc{MBHB} catalogue, assuming that HVSs are produced following the three-body interaction of a star with a MBHB. Here we also discuss the resulting number estimates. 
Finally, in Section \ref{sec:observations} we estimate \Gaia errors on the current sample of HVS candidates presented in \cite{brown+15}, and in Section \ref{sec:discussion} we summarize our results for the different catalogues, and we discuss their implications and limitations in view of the following data releases from the \Gaia satellite.

\section{The \textsc{"Vesc"} Mock Catalogue: A Simple Approach} 
\label{sec:mock}

We create synthetic populations of HVSs in order to assess and forecast \Gaia's performance in measuring their proper motions and parallax. We characterise the astrometric and photometric properties of the stars using their position in Galactic coordinates $(l,b,r)$ and mass $M$, and then estimate \Gaia's precision in measuring these properties.

In this section we choose to compute the total velocity $v$ of a HVS adopting a simple conservative approach, i.e. to assume it equal to the escape velocity from the Galaxy at its position:

\begin{equation}
\label{eq:vesc}
v(l,b,r) = v_\mathrm{esc}(l,b,r).
\end{equation}

Our decision is motivated by the choice not to focus on a particular ejection mechanism, but just to rely on the definition of a HVS as an unbound object. In addition to that, proper motions for a star travelling away from the GC on a radial orbit are directly proportional to the velocity, see equations \eqref{eq:mul} and \eqref{eq:mub}, therefore a higher velocity (e.g. for an unbound star) would result in a lower relative error in total proper motion, making the detection by \Gaia even more precise (refer to Section \ref{sec:mock_errors}). This catalogue does not make any assumption on the nature and origin of HVSs, and the impact of adopting a particular ejection mechanism for modelling the velocity distribution is explored in Sections \ref{sec:Hills} and \ref{sec:MBHB}, where we also introduce predictions for the expected bound population of HVSs.

For clarity and reference within this paper, we refer to this first catalogue as \textsc{Vesc}.

\subsection{Astrometric Characterization of a HVS} \label{sec:mock:astro}

In first approximation, HVSs are travelling away from the Milky Way on radial trajectories. This assumption holds if we consider the contribution given by the stellar disc to be sub-dominant in the total deceleration of the star \citep{kenyon+14}, and if we neglect deviations from spherical symmetry in the dark matter halo \citep{gnedin+05}. For a given position in the sky $(l,b,r)$, it is possible to derive the combination of proper motions in Galactic coordinates ($\mu_{l*} \equiv \mu_l \cos b, \ \mu_b)$ which is consistent with a star flying away from the GC on a straight line:
\begin{equation}
\label{eq:mul}
\mu_{l*}(l,b,r) = \frac{ \mathbf{\hat{p}} \cdot \mathbf{v}(l,b,r) }{r} = v(l,b,r)\frac{d_{\odot}}{r}\frac{\sin l}{r_\mathrm{GC}(l,b,r)}  ,
\end{equation}
\begin{equation}
\label{eq:mub}
\mu_{b}(l,b,r) = \frac{ \mathbf{\hat{q}} \cdot \mathbf{v}(l,b,r) }{r} = v(l,b,r)\frac{d_{\odot}}{r}\frac{\cos l \sin b}{r_\mathrm{GC}(l,b,r)} ,
\end{equation}
where $\mathbf{\hat{p}}$ and $\mathbf{\hat{q}}$ are unit basis vectors defining the plane tangential to the celestial sphere, $d_{\odot}$ is the distance between the Sun and the GC, and $r_{\mathrm{GC}}(l,b,r) = \sqrt{r^2 + d_\odot^2 - 2 r d_\odot \cos l \cos b}$ is the Galactocentric distance of the star. In the following, we will assume $d_\odot = 8.2$ kpc \citep{bland-hawthorn+16}. In order to simulate how these stars will appear in the \Gaia catalogue, we correct proper motions for the motion of the Sun and for the local standard of rest (LSR) velocity, following \cite{schonrich}.

The total velocity $v$, equal to the escape velocity from the Milky Way in that position, is computed assuming a three component Galactic potential: a Hernquist bulge \citep{hernquist90}:
\begin{equation}
\label{eq:bulge}
\phi_b(r_\mathrm{GC}) = -\frac{G M_b}{r_\mathrm{GC} + r_b},
\end{equation}
a Miyamoto \& Nagai disk in cylindrical coordinates $(R_\mathrm{GC}, z_\mathrm{GC})$ \citep{M&N75}:
\begin{equation}
\label{eq:disk}
\phi_d(R_\mathrm{GC}, z_\mathrm{GC}) = - \frac{GM_d}{\sqrt{R_\mathrm{GC}^2 + \Bigl(a_d + \sqrt{z_\mathrm{GC}^2 + b_d^2}\Bigr)^2}},
\end{equation}
and a Navarro-Frenk-White (NFW) halo profile \citep{nfw96}:
\begin{equation}
\label{eq:halo}
\phi(r_\mathrm{GC}) = -\frac{G M_h}{r_\mathrm{GC}} \ln\Bigl(1 + \frac{r_\mathrm{GC}}{r_s}\Bigr).
\end{equation}
The adopted values for the potential parameters $M_b$, $r_b$, $M_d$, $a_d$, $b_d$, $M_h$, and $r_s$ are summarized in Table \ref{Tab:potential}.
The mass and radius characteristic parameters for the bulge and the disk are taken from \cite{johnston+95, Price2014, hawkins+15}, while the NFW parameters are the best-fit values obtained in \cite{Rossi+17}. This choice of Galactic potential has been shown to reproduce the main features of the Galactic rotation curve up to $100$ kpc \citep[\cite{huang+16}, see Fig. A1 in][]{Rossi+17}.

\begin{table}
\centering 	
\caption{Parameters for the three-components Galactic potential adopted in the paper.}
\label{Tab:potential}
\begin{tabular}{l|l}
	\hline
	Component & Parameters \\
	\hline
	Bulge & $M_b = 3.4 \cdot 10^{10}$ M$_\odot$ \\
	 & $r_b = 0.7$ kpc \\
	Disk & $M_d = 1.0 \cdot 10^{11}$ M$_\odot$ \\
	 & $a_d = 6.5$ kpc \\
	 & $b_d = 0.26$ kpc \\
	Halo & $M_h = 7.6 \cdot 10^{11}$ M$_\odot$ \\
	 & $r_s = 24.8$ kpc \\
	\hline
\end{tabular}
\end{table}

As a result of \Gaia scanning strategy, the total number of observations per object depends on the ecliptic latitude of the star $\beta$, which we determine as \citep{jordi+10}:
\begin{equation}
\label{eq:ecliptic}
\sin \beta = 0.4971\sin b + 0.8677 \cos b \sin (l - 6.38^\circ).
\end{equation}

To complete the determination of the astrometric parameters, we simply compute parallax as $\varpi = 1/r$, where $\varpi$ is expressed in arcsec and $r$ in parsec.

\subsection{Photometric Characterization of a HVS} \label{sec:mock:photo}

Knowing the position and the velocity of a HVS in the Galaxy, we now want to characterize it from a photometric point of view, since \Gaia errors on the astrometry depend on the brightness of the source in the \Gaia passbands. 

To compute the apparent magnitudes in different bands, we need to know the age of the HVS at the given celestial location at the moment of its observation. This is required in order to correctly estimate its stellar parameters (radius, luminosity, and effective temperature) and the corresponding spectrum. We estimate the \emph{flight time} $t_\mathrm{f}$, the time needed to travel from the ejection region in the GC to the observed position, as:
\begin{equation}
\label{eq:tflight}
t_\mathrm{f}(l,b,r) = \frac{r_\mathrm{GC}(l,b,r)}{v_0(l,b,r)},
\end{equation}
where $v_0(r,l,b)$ is the velocity needed for a star in the GC to reach the observed position $(r,l,b)$ with zero velocity. We compute $v_0$ using energy conservation, evaluating the potential in the GC at $r=3$ pc, the radius of influence of the MBH \citep{genzel+10}. Since HVSs are decelerated by the Galactic potential, $t_\mathrm{f}$ is a lower limit on the actual flight time needed to travel from $3$ pc to the observed position. We then compare this time to the total main sequence (MS) lifetime $t_\mathrm{MS}(M)$, which we compute using analytic formulae presented in \cite{hurley+00}\footnote{We assume the MS lifetime to be equal to the total lifetime of a star.}, assuming a solar metallicity value \citep{brown15}. If $t_\mathrm{f}>t_\mathrm{MS}$ we exclude the star from the catalogue: its lifetime is not long enough to reach the corresponding position. On the other hand, if $t_\mathrm{f} < t_\mathrm{MS}$, we estimate the age of the star as:
\begin{equation}
\label{eq:tage}
t(M,l,b,r) = \varepsilon\bigl(t_\mathrm{MS}(M) - t_\mathrm{f}(l,b,r)\bigr),
\end{equation}
where $\varepsilon$ is a random number, uniformly distributed in $[0,1]$.

We evolve the star along its MS up to its age $t$  using analytic formulae presented in \cite{hurley+00}, which are functions of the mass and metallicity of the star. We are then able to get the radius of the star $R(t)$, the effective temperature $T_\mathrm{eff}(t)$, and the surface gravity $\log g(t)$. Chi-squared minimization of the stellar parameters $T_\mathrm{eff} (t)$ and $\log g (t)$ is then used to find the corresponding best-fitting stellar spectrum, and therefore the stellar flux, from the BaSeL SED Library $3.1$ \citep{westera+03}, assuming a mixing length of $0$ and a an atmospheric micro-turbulence velocity of $2$ \kms.

At each point of the sky we estimate the visual extinction $A_\mathrm{V}$ using the three-dimensional Galactic dust map \textsc{MWDUST}\footnote{\url{https://github.com/jobovy/mwdust}} \citep{bovy+16}. The visual extinction is then used to derive the extinction at other frequencies $A_\lambda$ using the analytical formulae in \cite{cardelli+89}, assuming $R_\mathrm{V} = 3.1$.

Given the flux $F(\lambda)$ of the HVS and the reddening we can then compute the magnitudes in the \Gaia $G$ band, integrating the flux in the \Gaia passband $S(\lambda)$ \citep{jordi+10}:
\begin{equation}
\label{eq:mag}
G = -2.5 \log \Biggl( \frac{ \int d\lambda \ F(\lambda) \ 10^{-0.4A_\lambda} \ S(\lambda)}{\int d\lambda \ F^{\mathrm{Vega}}(\lambda) \ S(\lambda)} \Biggr) + G^{\mathrm{Vega}}.
\end{equation}
The zero magnitude for a Vega-like star is taken from \cite{jordi+10}. Similarly, integrating the flux over the Johnson-Cousins $V$ and $I_C$ filters, we can compute the colour index $V - I_C$ \citep{bessel90}. We then compute the magnitude in the \Gaia $G_\mathrm{RVS}$ band using polynomial fits in \cite{jordi+10}.

\subsection{\emph{Gaia} Error Estimates} \label{sec:mock_errors}

We use the \textsc{Python} toolkit \textsc{PyGaia}\footnote{\url{https://github.com/agabrown/PyGaia}} to estimate post-commission, end-of-mission \Gaia errors on the astrometry of our mock HVSs. Measurement uncertainties depend on the ecliptic latitude, \Gaia $G$ band magnitude, and the $V-I_C$ colour of the star, which we all derived in the previous sections. We can therefore reconstruct \Gaia precision in measuring the astrometric properties of each HVS, which we quantify as the (uncorrelated) relative errors in total proper motion $z_{\mu} \equiv \sigma_\mu / \mu$, and in parallax $z_\varpi \equiv \sigma_\varpi / \varpi$. 

\subsection{Number Density of HVSs} \label{sec:density}

In order to determine how many HVSs \Gaia is going to observe with a given precision, we need to model their intrinsic number density. We assume a continuous and isotropic ejection from the GC at a rate $\dot{N}$. Indicating with $\rho(r_\mathrm{GC}, M)$ the number density of HVSs with mass $M$ at a Galactocentric distance $r_\mathrm{GC}$, we can simply write the total number of HVSs with mass $M$ within $r_\mathrm{GC}$ as:
\begin{equation}
\label{eq:NR}
N(<r_\mathrm{GC}, \ M) = \int_{0}^{r_\mathrm{GC}} 4\pi r^{\prime 2} \rho(r^\prime, \ M) dr^\prime.
\end{equation}
We assume HVSs to travel for a time $t_\mathrm{F}=r_\mathrm{GC}/v_\mathrm{F}$ to reach the observed position, where $v_\mathrm{F} = 1000$ km s$^{-1}$ is an effective average travel velocity. We also neglect the stellar lifetime after its MS, which could only extend by $\sim 10$\% the travel time. Current observations seem to suggest that the ejection of a HVS occurs at a random moment of its lifetime: $t_\mathrm{ej} = \eta t_\mathrm{MS}$ \citep{Brown+14}, with $\eta$ being a random number uniformly distributed in $[0, 1]$. We can then only observe a HVS at a distance $r_\mathrm{GC}$ if $t_\mathrm{F}$ satisfies:
\begin{equation}
\label{eq:tF}
t_\mathrm{F} = \frac{r_\mathrm{GC}}{v_\mathrm{F}} < t_\mathrm{MS} - t_\mathrm{ej} = t_\mathrm{MS}(1-\eta).
\end{equation}
We can then write the total number of HVSs of mass $M$ within $r_\mathrm{GC}$ as:
\begin{equation}
\label{eq:Nt}
N(<r_\mathrm{GC}, \ M) = \phi(M)\dot{N}\frac{r_\mathrm{GC}}{v_\mathrm{F}}\bigintsss_0^1 \theta\Biggl(t_\mathrm{MS}(1-\eta) - \frac{r_\mathrm{GC}}{v_\mathrm{F}} \Biggr) \ d\eta,
\end{equation}
where $\phi(M)$ is the mass function of HVSs, and $\theta(x)$ is the Heaviside step function. Differentiating this expression, we get:
\begin{equation}
\label{eq:dNdr}
\begin{split}
\frac{\partial N(<r_\mathrm{GC}, \ M)}{\partial r_\mathrm{GC}} = & \phi(M)\frac{\dot{N}}{v_\mathrm{F}}\bigintsss_0^1 \Biggl[\theta\Biggl(t_\mathrm{MS}(1-\eta)-\frac{r_\mathrm{GC}}{v_\mathrm{F}}\Biggr) \ + \\ & - \delta\Biggl(t_\mathrm{MS}(1-\eta)-\frac{r_\mathrm{GC}}{v_\mathrm{F}}\Biggr) \frac{r_\mathrm{GC}}{v_\mathrm{F}}  \Biggr] \ d\eta,
\end{split}
\end{equation}
where $\delta(x)$ is the Dirac delta function. Evaluating the integral and comparing this equation with the one obtained by differentiating equation \eqref{eq:NR} with respect to $r_\mathrm{GC}$, we can express the number density of HVSs within a given Galactocentric distance $r_\mathrm{GC}$ and with a given mass $M$ as:
\begin{equation}
\label{eq:density}
\begin{split}
\rho(r_\mathrm{GC},M) = & \theta\Biggl( t_{MS}(M) - \frac{r_\mathrm{GC}}{v_\mathrm{f}}\Biggr) \phi(M) \cdot  \Biggl( \frac{\dot{N}}{4\pi v_\mathrm{f} \ r_\mathrm{GC}^2} +  \\ &  -  \frac{\dot{N}}{2\pi r_\mathrm{GC} \ t_{MS}(M)v_\mathrm{f}^2} \Biggr) .
\end{split}
\end{equation}
\cite{Brown+14}, taking into account selection effects in the MMT HVS Survey, estimated a total of $\simeq 300$ HVSs in the mass range $[2.5$, $4]$ $M_\odot$ over the entire sky within $100$ kpc from the GC, that is:
\begin{equation}
\label{eq:N}
N\Bigl(r_\mathrm{GC}<100 \ \mathrm{kpc}, \ M\in [2.5, 4] \ M_\odot\Bigr) =  \varepsilon_\mathrm{f} \dot{N} \frac{ 100 \mathrm{kpc}}{v_\mathrm{f}}  = 300 .
\end{equation}
In this equation, $\varepsilon_\mathrm{f}$ is the mass fraction of HVSs in the $[2.5, 4]$ $M_\odot$ mass range, taking into account the finite lifetime of a star:
\begin{equation}
\label{eq:epsilonf}
\varepsilon_\mathrm{f} = \varepsilon_0 \bigintsss_{2.5 M_\odot}^{4 M_\odot} \phi(M) dM \bigintsss_{0}^{1}   \theta\Biggl( t_\mathrm{M}(1-\eta) - \frac{100 \mathrm{kpc}} {v_\mathrm{f}}  \Biggr) d\eta  .
\end{equation}
Assuming a particular mass function we can therefore estimate the ejection rate $\dot{N}$ needed to match observations using equation \eqref{eq:N} and \eqref{eq:epsilonf}. In the following we will assume a Kroupa IMF \citep{Kroupa01}, for which we get $\dot{N} \simeq 2.8 \cdot 10^{-4} \ \mathrm{year}^{-1}$. This estimate is consistent with other observational and theoretical estimates \citep{hills88, perets+07, zhang+13, Brown+14}.

For each object in the mock catalogue we can then compute the intrinsic number density of HVSs in that given volume $dV dM$ using equation \eqref{eq:density}. With a coordinate transformation to the heliocentric coordinate system, the corresponding number of HVSs in the volume element $dV \ dM$ is:
\begin{equation}
\label{eq:number}
\begin{split}
N(l,b,r,M) & = \rho(r_\mathrm{GC},M) dV dM \\ & = \rho(l,b,r,M) r^2 \cos b \ dl \ db \ dr \ dM.
\end{split}
\end{equation}

\subsection{\emph{"VESC"} Catalogue: Number Estimates of HVSs in \Gaia} 
\label{sec:estimates}

\begin{figure}
\centering
\includegraphics[width=0.5\textwidth]{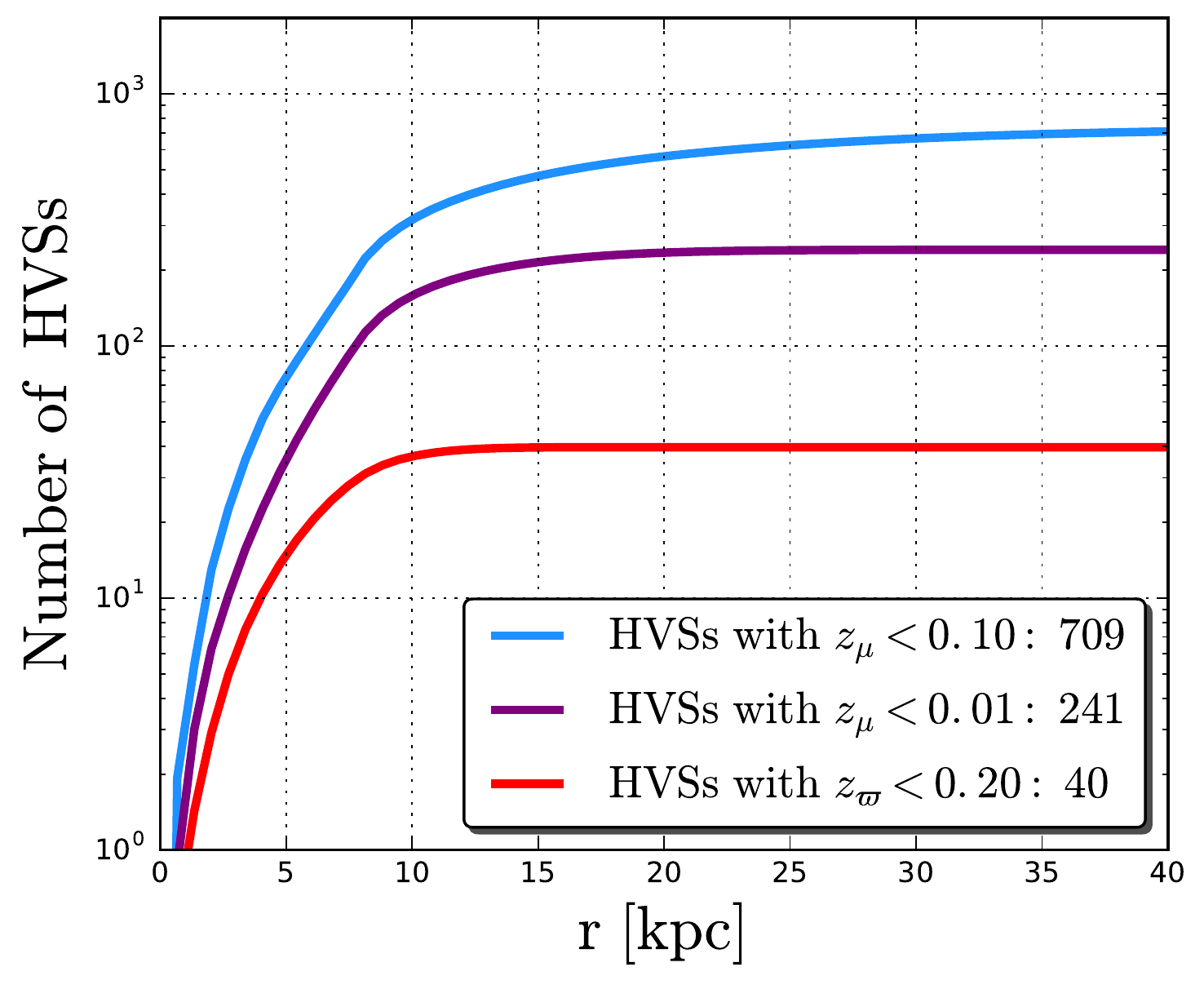}
\caption{\textsc{Vesc} catalogue: cumulative radial distributions of HVSs: the total number of HVSs within a heliocentric radius $r$. The blue (purple) line shows the cumulative radial distribution for HVSs which will be observable by \Gaia with a relative error on total proper motion below $10\%$ ($1\%$). The red line refers to those stars with a relative error on parallax below $20\%$.}
\label{fig:cumrad}
\end{figure}

We sample the space $(l,b,r,M)$ with a resolution of $\sim 6^\circ$ in $l$, $\sim 3^\circ$ in $b$, $\sim 0.7$ kpc in $r$ and $\sim 0.15$ $M_\odot$ in $M$. For each point we count how many HVSs lay in the volume element $dV \ dM$ using equation \eqref{eq:number}. We want to stress that the results refer to the end-of-mission performance of the \Gaia satellite.

Fig. \ref{fig:cumrad} shows the cumulative radial distribution of HVSs within $40$ kpc: stars which will be detectable by \Gaia with a relative error on total proper motion below $10\%$ ($1\%$) are shown with a blue (purple) line, and those with a relative error on parallax below $20\%$ with a red line. The total number of HVSs with a relative error on total proper motion below $10\%$ ($1\%$) is 709 (241). The total number of HVSs with a relative error on parallax below $20\%$ is $40$. We have chosen a relative error threshold of $0.2$ in parallax because, for such stars, it is possible to make a reasonable distance estimate by simply inverting the parallax, without the need of implementing a full Bayesian approach \citep{bailer-jones, astraatmadjaI, astraatmadjaII}. This is a great advantage, because uncertainties due to the distance determination dominate the errorbars in total velocity \citep{marchetti+17}. In all cases we can see that almost all detectable HVSs will be within 10 kpc from us.

\begin{figure}
\centering
\includegraphics[width=0.5\textwidth]{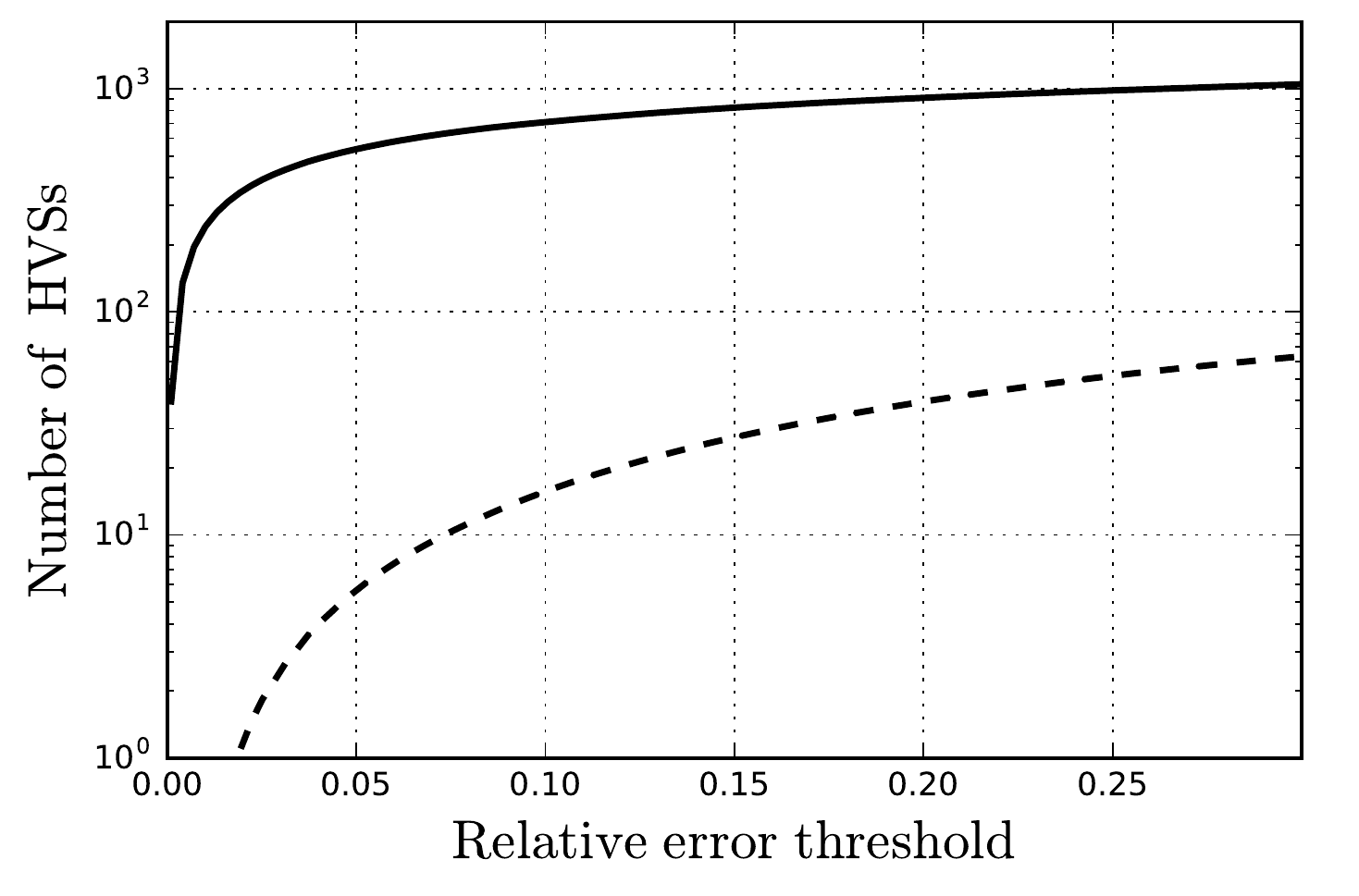}
\caption{\textsc{Vesc} catalogue: cumulative number of HVSs in the \Gaia catalogue for a relative error on total proper motion (solid line) and parallax (dashed line) within a given relative error threshold.}
\label{fig:f3}
\end{figure}

Fig. \ref{fig:f3} shows the total number of HVSs expected to be found in the \Gaia catalogue as a function of the chosen relative error threshold in total proper motion (solid) and parallax (dashed). We see that there is a total of $\sim 1000$ ($\sim 60$) HVSs with a relative error on total proper motion (parallax) below $30\%$. This imbalance reflects the lower precision with which \Gaia is going to measure parallaxes compared to proper motions.

\begin{figure}
\centering
\includegraphics[width=0.5\textwidth]{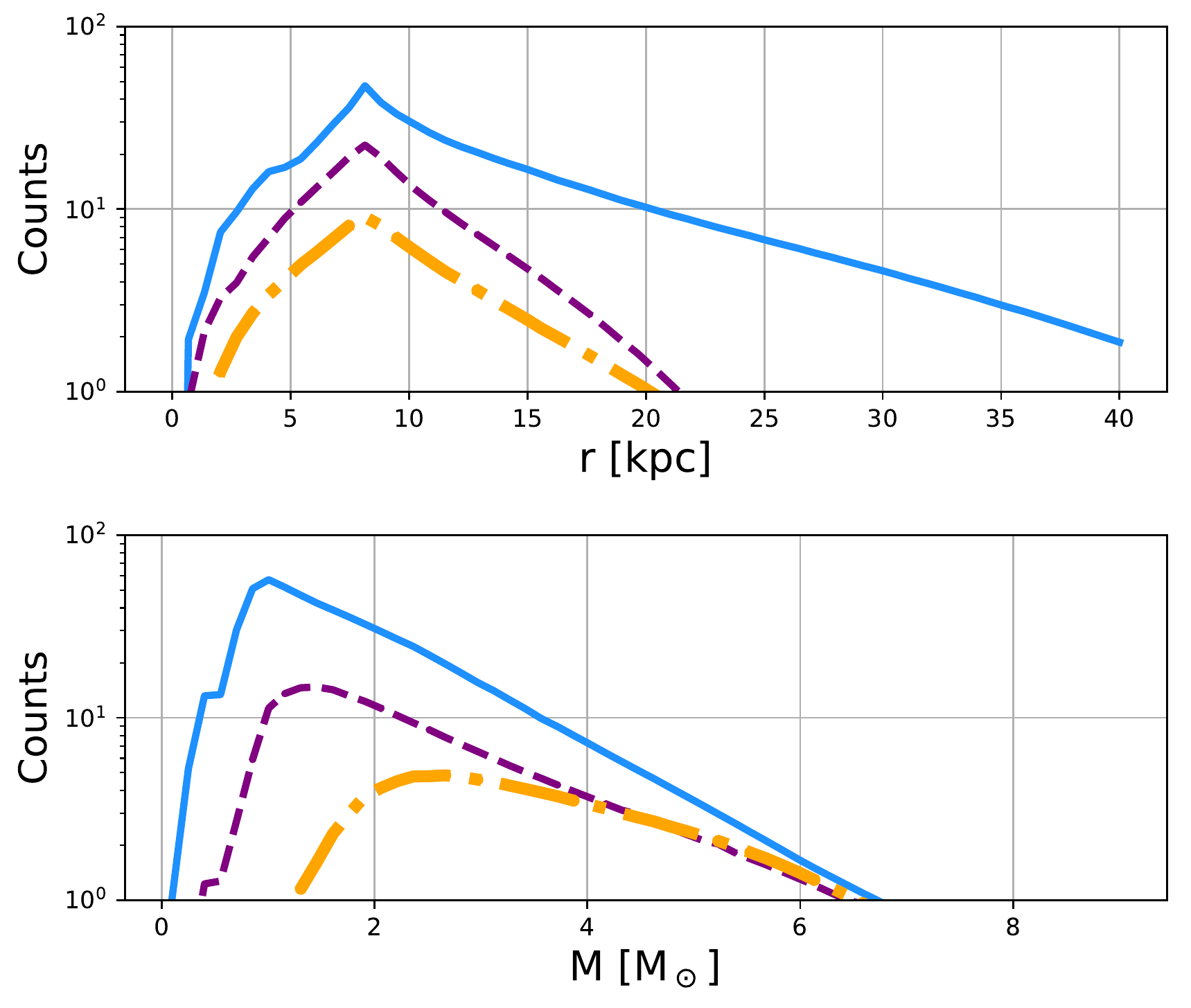}
\caption{\textsc{Vesc} catalogue: heliocentric distance (upper panel) and mass (lower panel) distribution for HVSs detectable by \Gaia with a relative error on total proper motion below $10 \%$ (solid), $1 \%$ (dashed), and for the golden sample of HVSs with a three-dimensional velocity by \Gaia alone (dot-dashed).} 
\label{fig:MRmu}
\end{figure}

Since proper motions are the most precise astrometric quantities, we quantify the radial and mass distribution of these  precisely-measured HVSs in Fig. \ref{fig:MRmu}. The solid and dashed curves refer, respectively, to stars detectable with a relative error on total proper motion below $10\%$ and $1\%$. Most HVSs with precise proper motions measurement will be at $r \simeq 8.5$ kpc, but the high-distance tail of the distribution extends up to $\sim 40$ kpc for HVSs with $z_\mu < 10\%$. The most precise proper motions will be available for stars within $\sim 20$ kpc from us. Also the mass distribution has a very well-defined peak which occurs at $M_\mathrm{peak} \simeq 1$ $M_\odot$, consistent with observational results in \cite{marchetti+17}. This is due to two main factors. The chosen IMF predicts many more low-mass than high mass stars, therefore we would expect a higher contribution from low-mass stars, but on the other hand low-mass stars tend to be fainter, and therefore will be detectable by \Gaia with a larger relative error. These two main contributions shape the expected mass function of HVSs in the catalogue.

\begin{figure}
\centering
\includegraphics[width=0.5\textwidth]{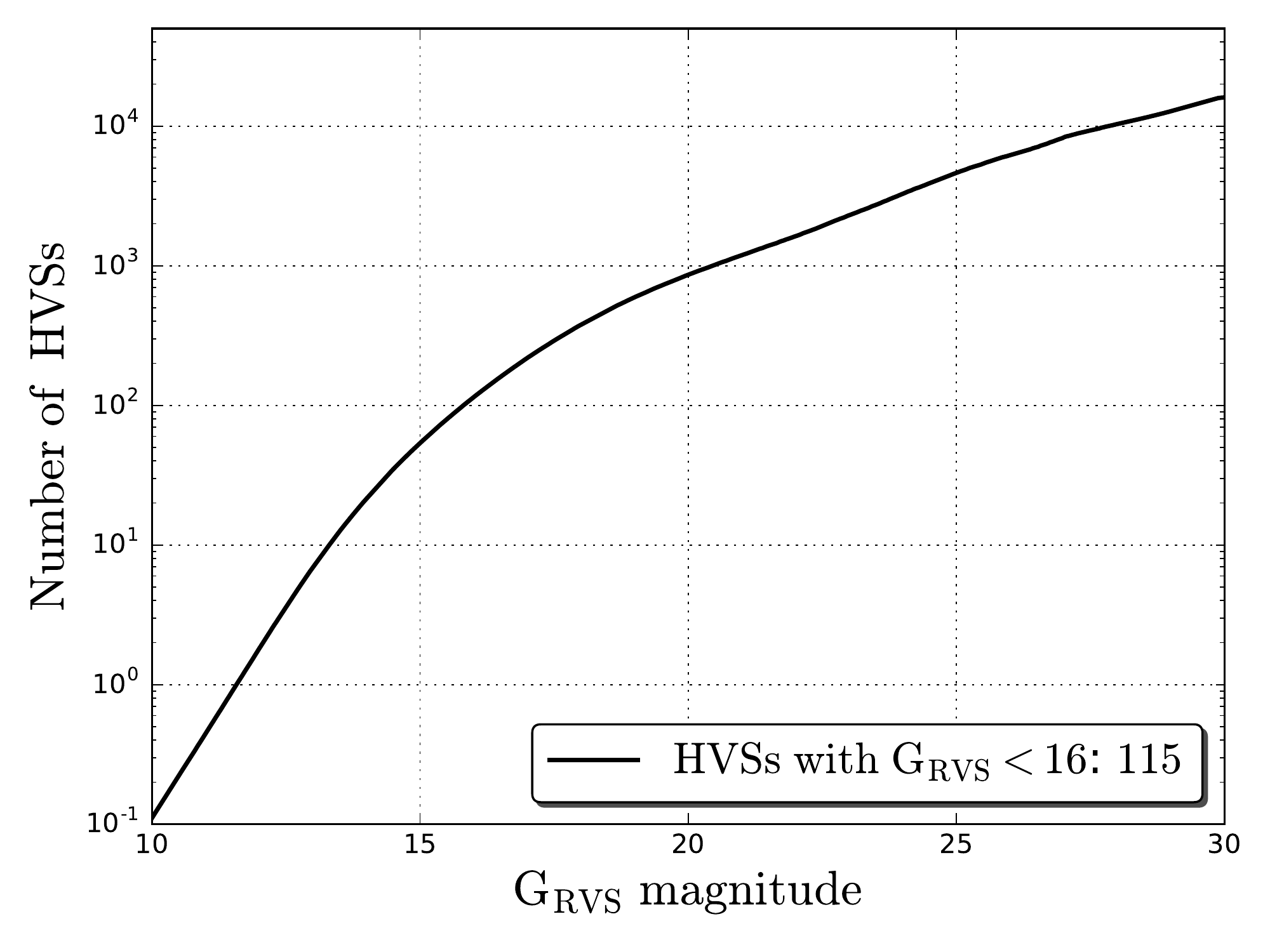}
\caption{\textsc{Vesc} catalogue: cumulative distribution of HVSs in the \Gaia G$_\mathrm{RVS}$ passband (the golden sample). We estimate a total of $115$ HVSs brighter than the $16$th magnitude in this filter.}
\label{fig:f5}
\end{figure}

Thanks to our mock populations and mock \Gaia observations, we can also determine for how many HVSs \Gaia will provide a radial velocity measurement. We refer to this sample as the $\emph{golden sample}$ of HVSs, since these stars will have a direct total velocity determination by \Gaia. To address this point we compute the cumulative distribution of magnitudes in the $G_\mathrm{RVS}$ passband, as shown in Fig. \ref{fig:f5}. There is a total of $115$ HVSs which satisfy the condition $G_\mathrm{RVS} < 16$, required for the Radial Velocity Spectrometer to provide radial velocities. The dot-dashed line in Fig. \ref{fig:MRmu} shows the distance and mass distribution for the golden sample of HVSs. The radial distribution is similar to the one shown in Fig.\ref{fig:MRmu}, with a peak at $r \simeq 8.5$ kpc. The mass distribution instead has a mean value $\simeq 3.6$ $M_\odot$ and a high-mass tail which extends up to $\simeq 6$ $M_\odot$.

\begin{figure}
\centering
\includegraphics[width=0.5\textwidth]{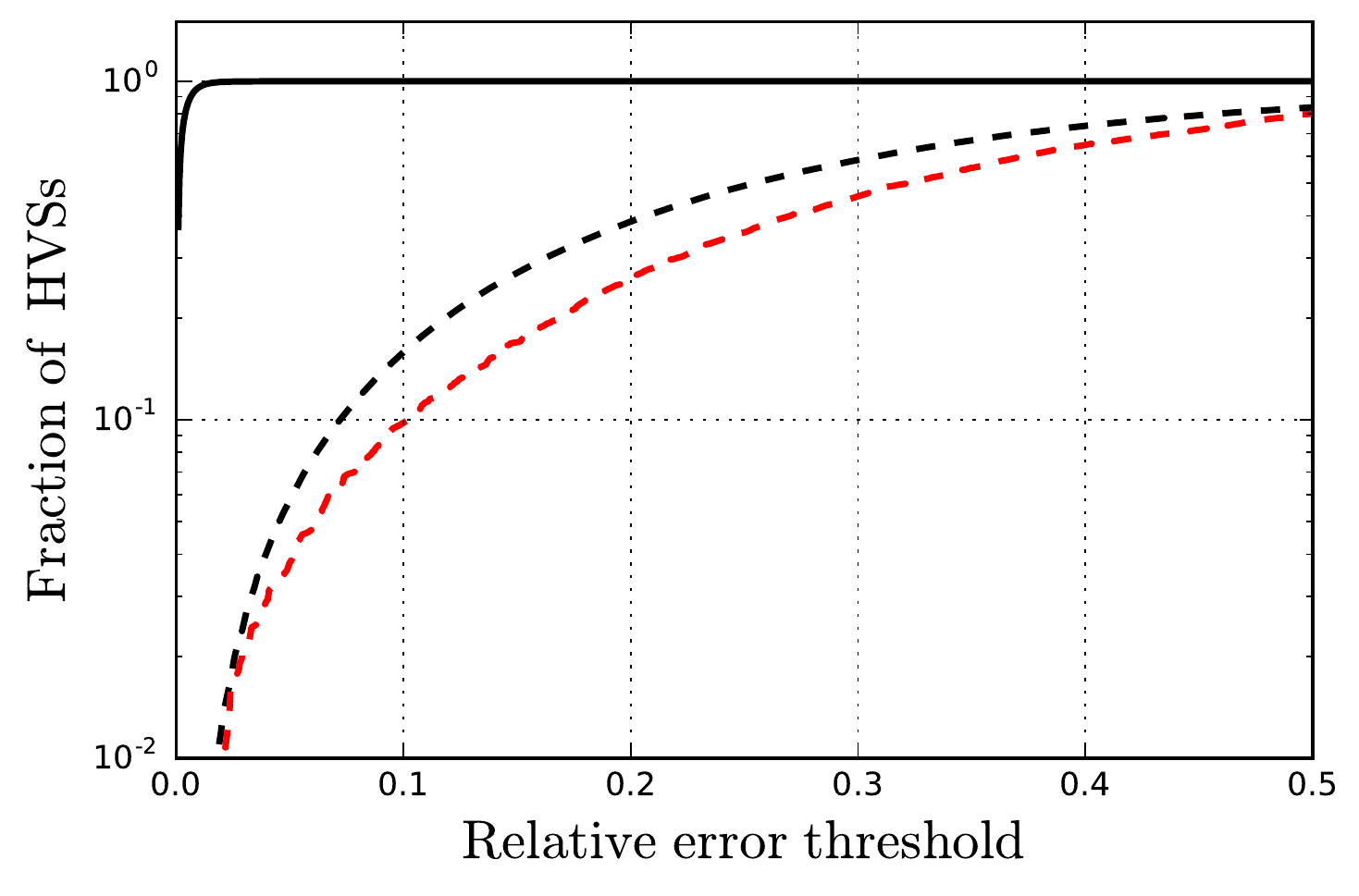}
\caption{Cumulative fraction of HVSs in the golden sample within a certain threshold for relative errors in total proper motion (solid) and parallax (dashed). The black curves refer to the \textsc{Vesc} catalogue, while the red dashed one to the \textsc{Hills} catalogue (refer to Section \ref{sec:Hills}). The red solid line overlaps with the black one, therefore it is not shown in the plot. The two curves for the \textsc{MBHB} catalogue coincide with the ones for the \textsc{Vesc} catalogue, and thus are not shown.}
\label{fig:err_bright}
\end{figure}

Fig. \ref{fig:err_bright} shows the cumulative distribution function of stars in the golden sample with a relative error on proper motion (solid) and on parallax (dashed) below a given threshold. This plot shows that proper motions will be detected with great accuracy for all of the stars: $z_\mu \lesssim 0.4\%$ over the whole mass range. $39$ of these stars ($34 \%$ of the whole golden sample) will have $z_\varpi < 20\%$, and therefore it will be trivial to determine a distance for these stars, by simply inverting the parallax.

\subsubsection{Estimates in \Gaia DR1/TGAS and DR2}

On September 14th $2016$, \Gaia DR1 provided positions and $G$ magnitudes for all sources with acceptable errors on position ($1142679769$ sources), and the full five-parameters solution ($\alpha$, $\delta$, $\varpi$, $\mu_{\alpha*}$, $\mu_\delta$) for stars in common between \Gaia and the \emph{Tycho}-$2$ catalogue ($2057050$ sources, the TGAS catalogue) \citep{gaiaa, gaiab, lindegren+16}.

To estimate the number of HVSs expected to be found in the TGAS subset of the first data release, we repeat the analysis of Section \ref{sec:estimates} considering the principal characteristics of the \emph{Tycho}-$2$ star catalogue \citep{hog+00}. We employ a $V < 11$ magnitude cut, corresponding to the $\sim 99\%$ completeness of the \emph{Tycho}-$2$ catalogue \citep{hog+00}. We find a total of $0.46$ HVSs surviving this magnitude cut. This result is consistent with results in \cite{marchetti+17}, which find only one star with both a predicted probability $> 50\%$ of being unbound from the Galaxy and a trajectory consistent with coming from the GC.

\Gaia data release 2, planned for April 2018, will be the first release providing radial velocities. It will consists of the five-parameter astrometric solution for the full billion star catalogue, and radial velocity will be provided for stars brighter than $G_\mathrm{RVS} = 12$. We find a total of $2$ HVSs to survive the $G_\mathrm{RVS} < 12$ magnitude cut. 

\section{The \textsc{"Hills"} Catalogue}  \label{sec:Hills}

In the previous analysis we derived model independent estimates for unbound stars, by assuming that the total velocity of a HVS in a given point is equal to the local escape velocity from the Milky Way. In this and the next section, we instead employ a physically motivated velocity distribution. In this section we adopt the Hills mechanism \citep{hills88}, the most successful ejection mechanism for explaining current observations \citep{brown15}. In this case we will have a population of bound HVSs, in addition to the unbound ones (see discussion in Section \ref{sec:intro}). We call this catalogue \textsc{Hills}, to differentiate it from the simpler \textsc{Vesc} catalogue introduced and discussed in Section \ref{sec:mock}.

\subsection{Velocity Distribution of HVSs}

We start by creating a synthetic population of binaries in the GC, following and expanding the method outlined in \cite{Rossi+17} and \cite{marchetti+17}. We identify three parameters to describe binary stars: the mass of the primary $m_p$ (the more massive star), the mass ratio between the primary and the secondary $q < 1$, and the semi-major axis of the orbit $a$. For the primary mass, we assume a Kroupa initial mass function in the range $[0.1, 100]$ M$_\odot$, which has been found to be consistent with the initial mass function of stellar populations in the GC \citep{bartko+10}. We assume power-laws for the distributions of mass ratios and semi-major axes: $f_q \propto q^\gamma$, $f_a \propto a^\alpha$, with $\gamma =-1$, $\alpha = -3.5$. This combination is consistent with observations of B-type binaries in the $30$ Doradus star forming region of the LMC \citep{dunstall+15}, and provides a good fit to the known HVS candidates from the HVS survey for reasonable choices of the Galactic potential \citep{Rossi+17}. The lower limit for $a$ is set by the Roche lobe overflow: $a_\mathrm{min} = 2.5 \max(R_p, R_s)$, where $R_p$ and $R_s$ are, respectively, the radius of the primary and secondary star. The radius is approximated using the simple scaling relation $R_ i\propto m_i$, with $i = p, s$. We arbitrarily set the upper limit of $a$ to $2000$ R$_\odot$.

\cite{kobayashi+12} showed that, for a binary approaching the MBH on a parabolic orbit, there is an equal probability of ejecting either the primary or the secondary star in the binary. We then randomly label one star per binary as HVS (mass $M$) and the other one as the bound companion (mass $m_c$). Following \cite{sari+10, kobayashi+12, rossi+14} we then sample velocities from an ejection distribution which depends analytically on the properties of the binary approaching the MBH:
\begin{equation}
\label{eq:vdistr}
v_\mathrm{ej} = \sqrt{ \frac{2Gm_\mathrm{c}}{a}} \Biggl(\frac{M_\bullet}{m_\mathrm{t}}\Biggr)^{1/6},
\end{equation}
where $M_\bullet = 4.3 \cdot 10^6$ $M_\odot$ is the mass of the MBH in our Galaxy \citep{ghez+08, gillessen+09, meyer+12}, $m_\mathrm{t} = M + m_\mathrm{c}$ is the total mass of the binary, and $G$ is the gravitational constant. This equation represents the resulting ejection velocity after the disruption of the binary for a star at infinity with respect only to the MBH potential. Rigorously, there should be a numerical factor depending on the geometry of the three-body encounter in front of the square root, but it has been shown to be of the order of unity when averaged over the binary phase, and not to influence the overall velocity distribution \citep{sari+10, rossi+14}.

\subsection{Flight Time Distribution of HVSs}

\begin{figure}
	\centering
	\includegraphics[width=0.35\textwidth]{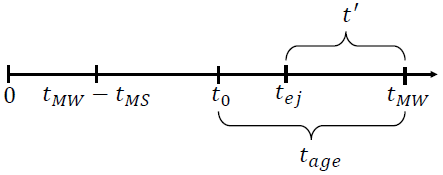}
	\caption{Sequence of events in the life of a HVS with a total lifetime $t_\mathrm{MS}(M) < t_\mathrm{MW}$. The instant $0$ corresponds to the time when the MW was formed, while $t_\mathrm{MW}$ is today, when we observe the HVS in the sky. The time $t_0$ ($t_\mathrm{ej}$) is the age of the Galaxy when the HVS was born (ejected). The time $t^\prime$ is the flight time of the HVS, while $t_\mathrm{age}$ is its present age.}
	\label{fig:timeline}
\end{figure}

Following the discussion in Section \ref{sec:mock:photo}, the flight time $t'$ of a HVS is defined as the time between its ejection from the GC and its observation. We assume the total lifetime of a star of mass $M$ to be equal to its main sequence lifetime $t_\mathrm{MS}(M)$, and we also assume $t_\mathrm{MW} = 13.8$ Gyr to be the current age of the MW \citep{planck16}. We compute the average fight time for stars to which the condition $t_\mathrm{MS}(M) < t_\mathrm{MW}$ applies. We call $t_0$ and $t_\mathrm{ej}$, respectively, the age of the Galaxy at the instant when a HVS visible today is born and when the star is ejected. We assume $t_0$ to be distributed uniformly between $t_\mathrm{MW} - t_\mathrm{MS}(M)$ and $t_\mathrm{MW}$:
\begin{equation}
t_0(M) = t_\mathrm{MW}  - t_\mathrm{MS}(M)(1 - \epsilon_1),
\end{equation}
and $t_\mathrm{ej}$ to be distributed uniformly between $t_0(M)$ and $t_\mathrm{MW}$:
\begin{equation}
t_\mathrm{ej}(M) = t_0(M) + \epsilon_2(t_\mathrm{MW} - t_0(M)).
\end{equation}
In the above expressions, $\epsilon_1$ and $\epsilon_2$ are two random numbers uniformly distributed in $[0, 1]$. Finally, we can express the flight time of a HVS as:
\begin{equation}
\label{eq:tf}
t^\prime(M) = t_\mathrm{MW} - t_\mathrm{ej}(M) = \varepsilon_1 \varepsilon_2 t_\mathrm{MS}(M),
\end{equation}
where $\varepsilon_1 \equiv (1 - \epsilon_1)$ and $\varepsilon_2 \equiv (1 - \epsilon_2)$ are two random numbers uniformly distributed in $[0, 1]$. Figure \ref{fig:timeline} visually presents the relevant time intervals. The probability density function for $t^\prime$ is then:
\begin{equation}
\label{eq:tfPDF}
f(t^\prime, M) = -\frac{1}{t_\mathrm{MS}(M)} \log \frac{t^\prime(M)}{t_\mathrm{MS}(M)}.
\end{equation}
We can then write the survival function $g(t^\prime, M)$, the fraction of HVSs alive at a time $t^\prime$ after the ejection, as:
\begin{equation}
\label{eq:gt'}
g(t^\prime, M) = 1 - \int_0^{t^\prime}f(\tau, M) d\tau = 1 + \frac{t^\prime(M)}{t_\mathrm{MS}(M)} \Biggl(\log\frac{t^\prime(M)}{t_\mathrm{MS}(M)} - 1 \Biggr).
\end{equation}
We can express the age of a HVS at the moment of its observation as:
\begin{equation}
\label{eq:tageH}
t_\mathrm{age}(M) = t_\mathrm{MW} - t_0(M) = \varepsilon_1 t_\mathrm{MS}(M).
\end{equation}
To take into account low-mass stars with $t_\mathrm{MS}(M) \ge t_\mathrm{MW}$, we rewrite equations \eqref{eq:tf} and \eqref{eq:tageH} as:
\begin{equation}
t^\prime(M) = 
	\begin{cases}
	\varepsilon_1 \varepsilon_2 t_\mathrm{MS}(M) & \mathrm{if \ } t_\mathrm{MS}(M) < t_\mathrm{MW} \\
	\varepsilon_1 \varepsilon_2 t_\mathrm{MW} & \mathrm{if \ } t_\mathrm{MS}(M) \ge t_\mathrm{MW} 
	\end{cases},
\end{equation}
\begin{equation}
\label{eq:tageH_MW}
t_\mathrm{age}(M) = 
	\begin{cases}
	\varepsilon_1 t_\mathrm{MS}(M) & \mathrm{if \ } t_\mathrm{MS}(M) < t_\mathrm{MW} \\
	\varepsilon_1 t_\mathrm{MW} & \mathrm{if \ } t_\mathrm{MS}(M) \ge t_\mathrm{MW}
	\end{cases}.
\end{equation}

\subsection{Initial Conditions and Orbit Integration} \label{sec:orbits}

The ejection velocity for the Hills mechanism, given by equation \eqref{eq:vdistr}, is the asymptotic velocity of a HVS at an infinite distance from the MBH. In practice, we model this distance as the radius of the gravitational sphere of the influence of the black hole, which is constrained to be of the order of $\bar{r}_0 = 3$ pc \citep{genzel+10}.  
We then initialize the position of each star at a distance of $\bar{r}_0$, with random angles (latitude, longitude) drawn from uniform spherical distributions. Velocities are drawn according to equation \eqref{eq:vdistr}, and the velocity vector is chosen is such a way to point radially away from the GC at the given initial position, so that the angular momentum of the ejected star is zero.

The following step is to propagate the star in the Galactic potential up to its position $(l,b,r)$ after a time $t'$ from the ejection. We do that assuming the potential model introduced in Section \ref{sec:mock:astro}. The orbits are integrated using the publicly available \textsc{Python} package \textsc{galpy}\footnote{\url{https://github.com/jobovy/galpy}} \citep{bovy15} using a Dormand-Prince integrator \citep{dormand&prince80}. The time resolution is kept fixed at $0.015$ Myr. We check for energy conservation as a test for the accuracy of the orbit integration.  

We therefore obtain for each star its total velocity $v$ in the observed position, and we build a mock catalogue of HVSs with relative errors on astrometric properties, following the procedure outlined in Sections \ref{sec:mock:astro} to \ref{sec:mock_errors}.

\subsection{\textsc{"HILLS"} Catalogue: Number Estimates of HVSs in \Gaia}
\label{sec:Hillsestimates}

\begin{figure}
	\centering
	\includegraphics[width=0.5\textwidth]{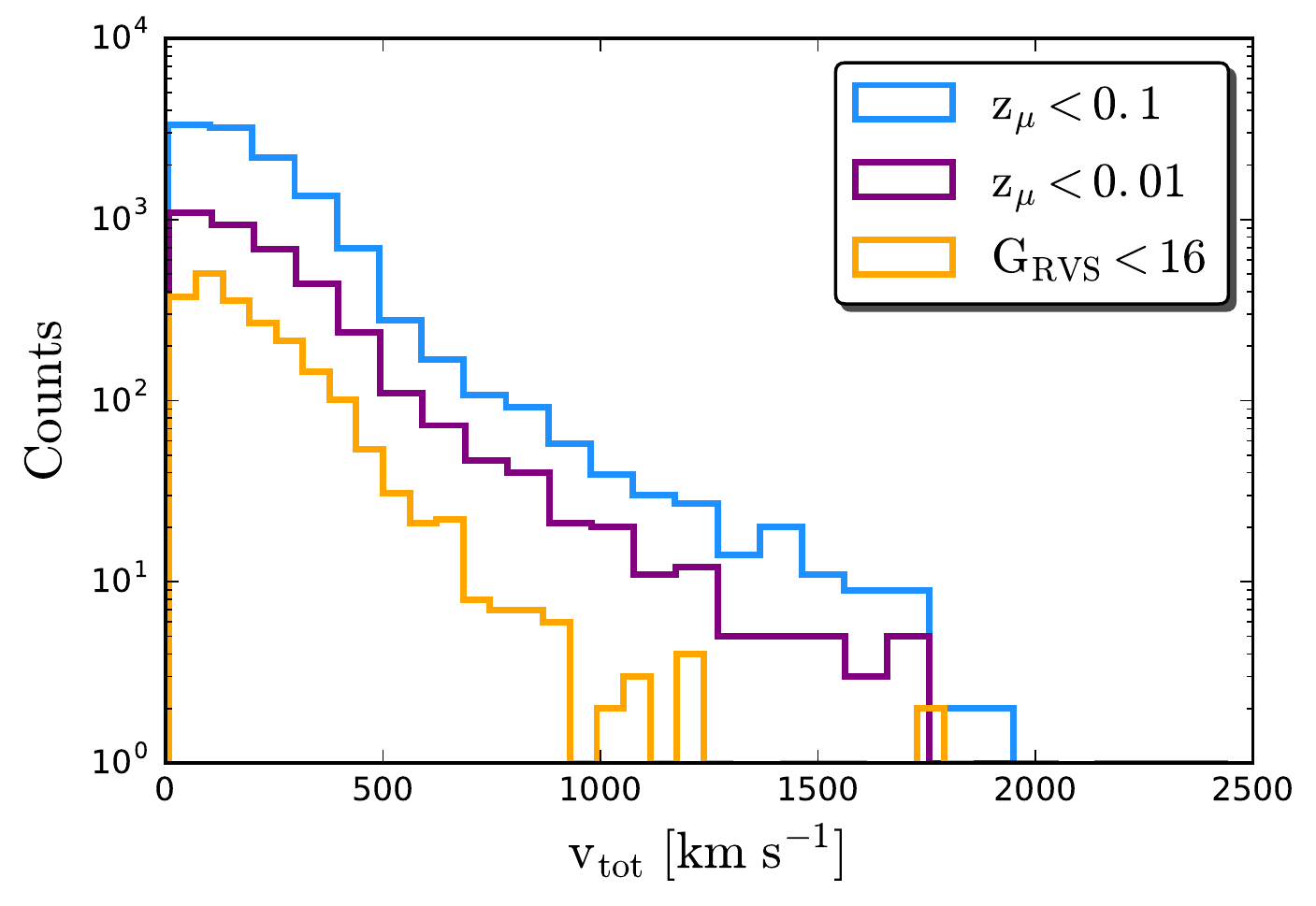}
	\caption{\textsc{Hills} catalogue: distribution of total velocities in the Galactocentric rest frame for the HVSs with a relative error on total proper motion below 10$\%$ (blue), 1$\%$ (purple), and with a radial velocity measurement (yellow).}
	\label{fig:vtotdistr}
\end{figure}

\begin{figure}
	\centering
	\includegraphics[width=0.5\textwidth]{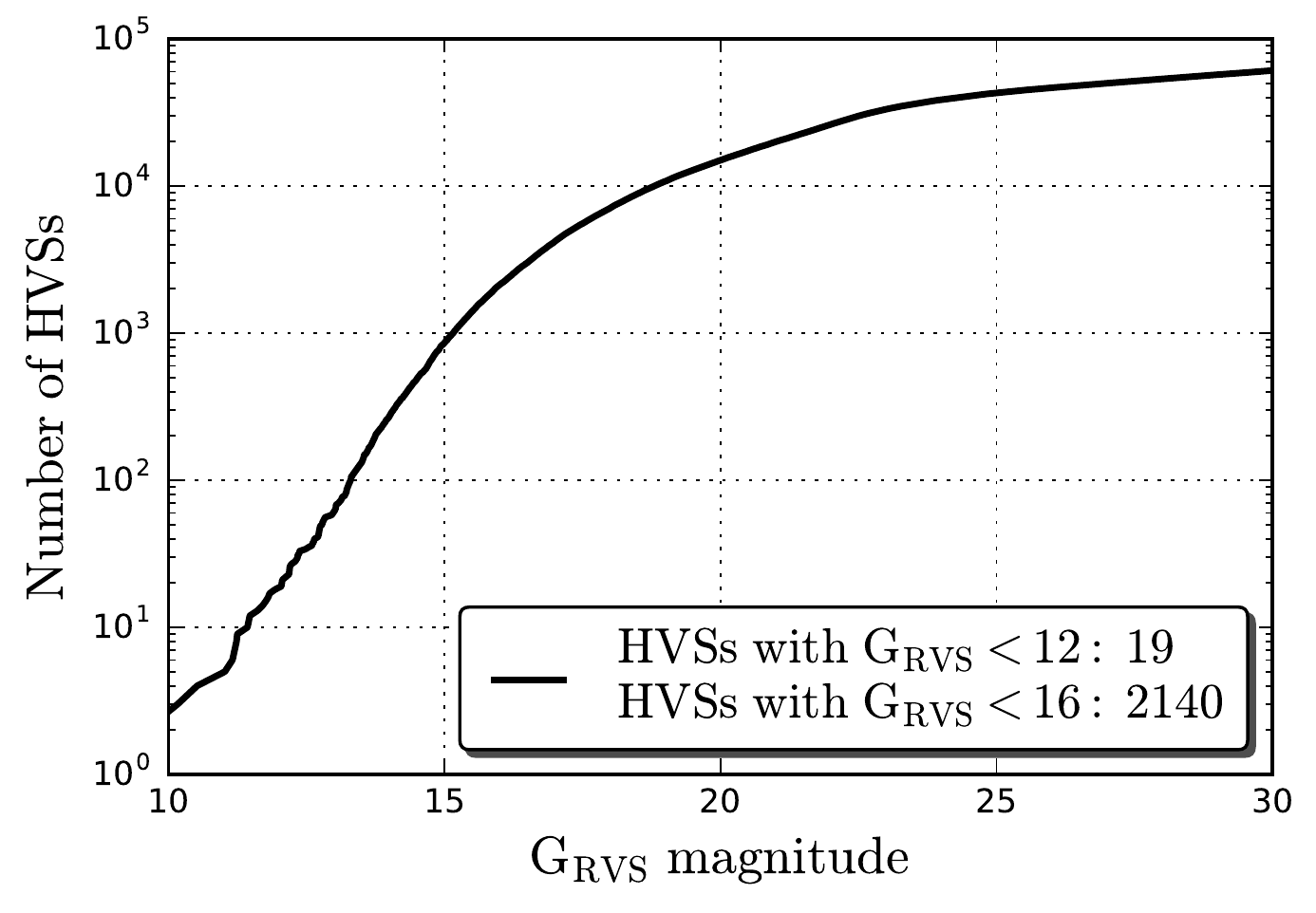}
	\caption{\textsc{Hills} catalogue: cumulative distribution of HVSs in the \Gaia G$_\mathrm{RVS}$ passband. We estimate a total of $2140$ HVSs brighter than the $16$th magnitude in this filter, and $19$ HVSs brighter than the $12$th magnitude.}
	\label{fig:Gdistr}
\end{figure}

\begin{figure*}
	\centering
	\includegraphics[width=0.75\textwidth]{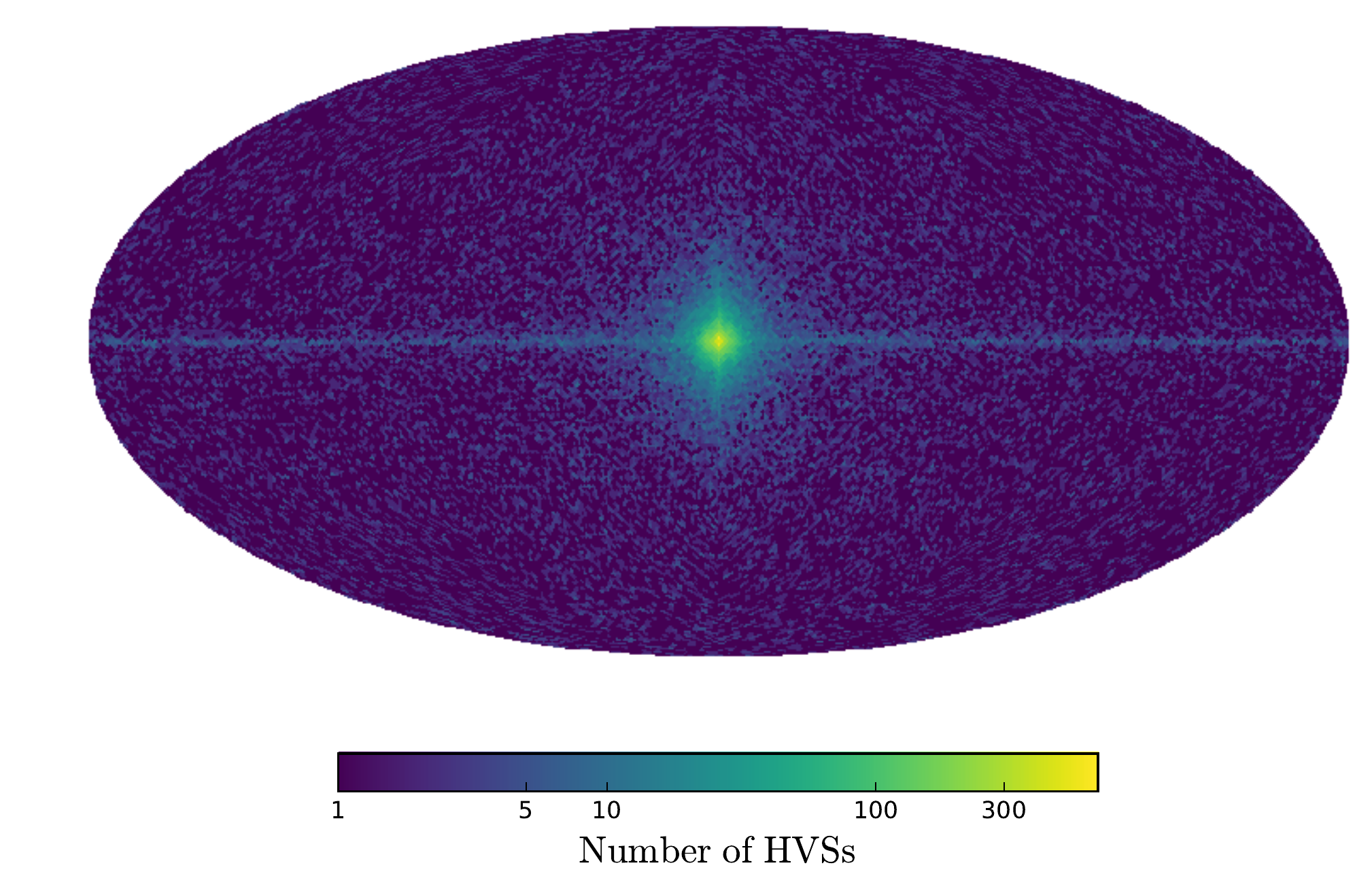}
	\caption{\textsc{Hills} catalogue: sky distribution in Galactic coordinates of the current population of HVSs in our Galaxy ($10^5$ stars).}
	\label{fig:allHVSs}
\end{figure*}

\begin{figure*}
	\centering
	\includegraphics[width=0.8\textwidth]{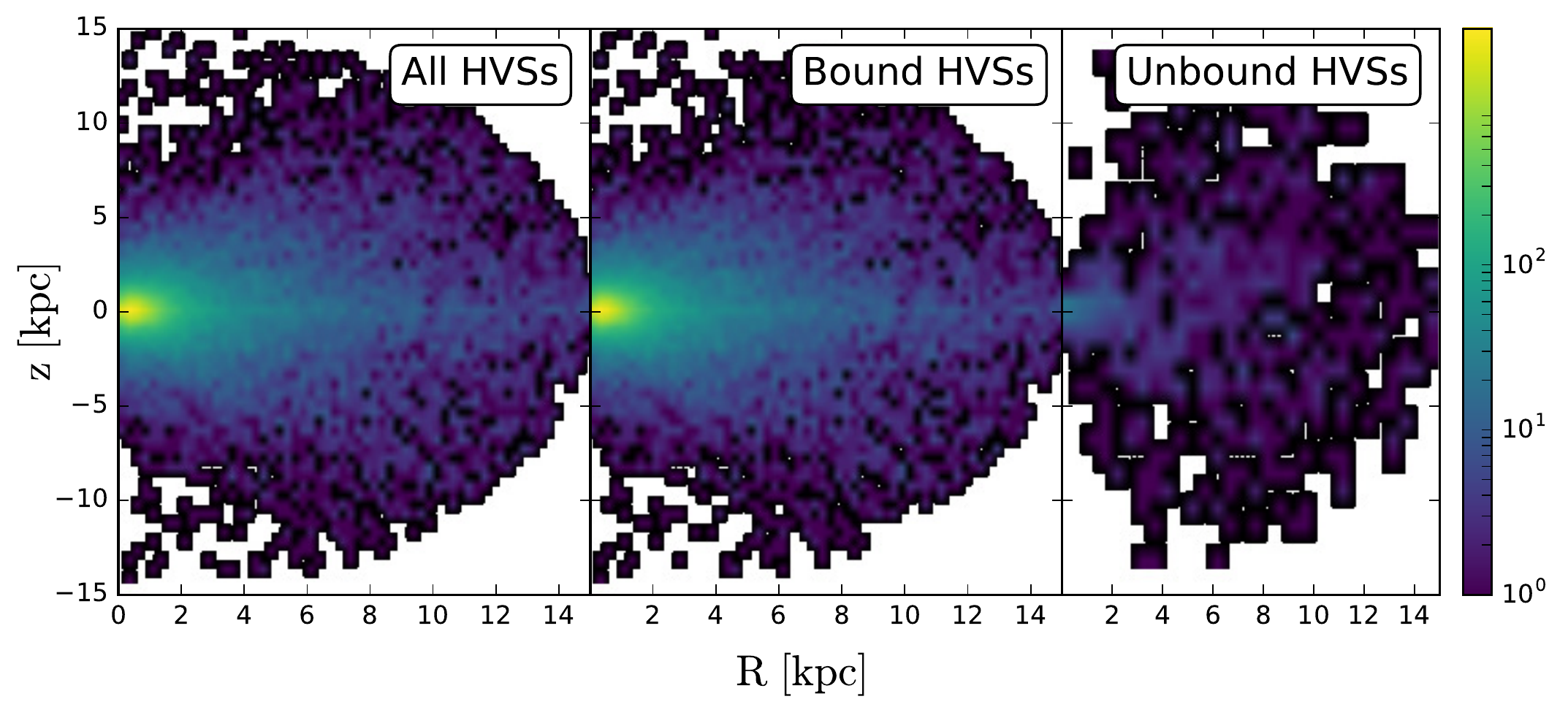}
	\caption{\textsc{Hills} catalogue: distribution in Galactocentric cylindrical coordinates $(R, z)$ of all HVSs (left), bound HVSs (centre), and unbound HVSs (right) within $15$ kpc from the Galactic Centre.}
	\label{fig:Rz}
\end{figure*}

We start by estimating the number of HVSs currently present in our Galaxy. We call $\frac{dn}{dM}(M)$ the normalized probability density function of masses upon ejection. We note that this is not a Kroupa function, because the HVS is not always the primary star of the binary, and the secondary star is drawn according the mass ratio distribution $f_q \propto q^{-3.5}$. Assuming that HVSs have been created at a constant rate $\eta$ for the entire Milky Way's lifetime $t_\mathrm{MW}$, the present Galactic population of HVSs in the mass range $[0.5, 9]$ M$_\odot$ is:
\begin{equation}
\label{eq:Ntot}
N = \eta \int_o^{t_\mathrm{MW}} dt^\prime \int_{0.5 \mathrm{M}_\odot}^{9 \mathrm{M}_\odot} dM \frac{dn}{dM}(M) g(t^\prime, M).
\end{equation}
We choose to restrict ourselves to the mass range $[0.5, 9]$ M$_\odot$ because stars with higher or lower masses are, respectively, very rare given our chosen IMF or not bright enough to be detectable by \Gaia with good precision. Assuming the value $\eta = 2.8 \cdot 10^{-4}$ yr$^{-1}$ derived in Section \ref{sec:density}, anchored to the current observations of HVSs, we get $N \simeq 10^5$. We thus generate $10^5$ HVSs in the GC as explained in the previous sections, and we propagate them in the Galaxy.

We can now use this realistic mock catalogue to predict the main properties of the Galactic population of HVSs. We find:

\begin{itemize}

	\item $52 \%$ of the total number of stars travel along unbound orbits. Note that this does \emph{not} imply that most of the HVSs will be detected with high velocities: given our choice of the Galactic potential, the escape velocity curve decreases to a few hundreds of km s$^{-1}$ at large distances from the GC ($\gtrsim 100$ kpc). Therefore a large number of HVSs is classified as unbound even if velocities are relatively low. In particular, we find $5\%$ ($6$\%) of the stars with $z_\mu < 0.1$ ($z_\mu < 0.01$) to be unbound from the MW. The distribution of total velocities in the Galactic rest frame is shown in Fig. \ref{fig:vtotdistr}, where we can see that the distribution peaks at $v < 500$ \kms. The blue (purple) curve refers to HVSs that will be detected by \Gaia with a relative error on total proper motion below $10\%$ ($1\%$), while the yellow curve is the distribution of HVSs with a radial velocity measurement. We can see that majority of stars with extremely high velocities ($v \gtrsim 1000$ \kms) will not be brighter than $G_\mathrm{RVS}= 16$, but few of them will be included in the catalogue, becoming the fastest known HVSs. The majority of stars, having low velocities, could easily be mistaken for disc, halo, or runaway stars, based on the module of the total velocity only (refer to discussion in Section \ref{sec:discussion}).
	
	\item $2.1 \%$ of the HVSs will have G$_\mathrm{RVS} < 16$ with \Gaia radial velocities. This amounts to $2140$ stars. The proper motion and parallax error distributions for this golden sample of HVSs are shown in Fig. \ref{fig:err_bright}. The cumulative distribution function of $G_\mathrm{RVS}$ magnitudes for all stars in the mock catalogue is shown in Fig. \ref{fig:Gdistr}. $68$ of the G$_\mathrm{RVS} < 16$ stars are unbound. $165$ of the G$_\mathrm{RVS} < 16$ have total velocity above $450$ km s$^{-1}$.
	
	\item From Fig. \ref{fig:Gdistr} we can see that $19$ stars are brighter than the $12$th magnitude in the G$_\mathrm{RVS}$ band, so there will be direct \Gaia radial velocity measurements already in \Gaia DR2. We find $0$ of these stars to be unbound from the MW. Proper motion error estimates for \Gaia DR2 can be obtained rescaling the errors from \textsc{PyGaia} by a factor\footnote{This numerical factor is derived considering that \Gaia DR2 uses $21$ months of input data, and that the error on proper motion scales as $t^{-1.5}$ (taking into account both the photon noise and the limited time baseline).} $(60/21)^{1.5} \sim 4.8$. We find all the $19$ stars to have relative errors in total proper motion $\lesssim 0.01\%$, and in parallax $\lesssim 20\%$. 
	
	\item $250$ unbound HVSs with masses in $[2.5, 4]$ M$_\odot$ are within $100$ kpc from the GC. This number is consistent with the observational estimate in \cite{Brown+14}.
\end{itemize}

Fig. \ref{fig:allHVSs} shows the distribution in Galactic coordinates of the population of $10^5$ HVSs, while Fig. \ref{fig:Rz} shows the distribution in Galactocentric cylindrical coordinates of the HVSs within $15$ kpc from the Galactic Centre. In all cases we can see that most HVSs lie in the direction of the GC: $(l,b) = (0,0)$. This is due to the presence of the population of bound HVSs, whose velocity is not high enough to fly away from the Milky Way, and therefore they spend their lifetime in the central region of the Galaxy on periodic orbits. Fig. \ref{fig:Rz} also shows how the majority of HVSs in the inner part of the Galaxy are travelling on bound orbits.

\begin{figure}
	\centering
	\includegraphics[width=0.5\textwidth]{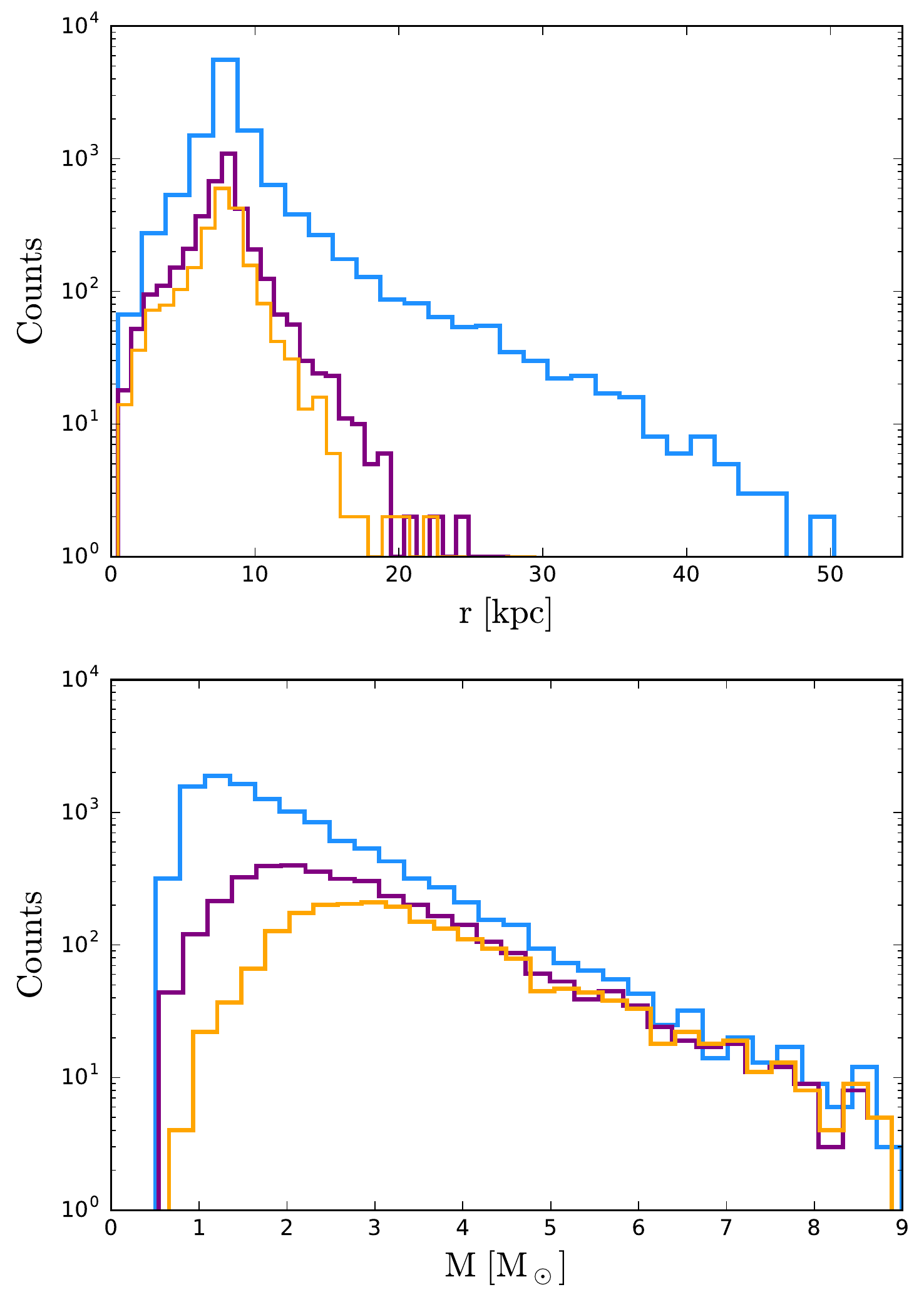}
	\caption{\textsc{Hills} catalogue: heliocentric distance (top) and mass (bottom) distribution of the HVSs with a relative error on total proper motion below 10$\%$ (blue), 1$\%$ (purple), and with a radial velocity measurement by \Gaia (yellow).}
	\label{fig:Mrdistr}
\end{figure}

The distance distribution of the HVS sample is shown in the top panel of Fig. \ref{fig:Mrdistr} for three samples: stars with a relative error on total proper motion below $10\%$ (blue), below $1\%$ (purple), and with a three-dimensional velocity determination (yellow). We can see that most stars lie within few tens of kpc from us, with only a few objects at distances $\sim 50$ kpc. We also note the substantial overlap between the purple and the yellow histogram, suggesting again that HVS with a radial velocity measurement will have an accurate \emph{total} velocity by \Gaia. The peak in the distributions, below $10$ kpc, well agrees with the one shown in Fig. \ref{fig:MRmu}.

We show the mass distribution of the sample of HVSs in the bottom panel of Fig. \ref{fig:Mrdistr}. The colour code is the same as before. As expected, massive stars are brighter, and will therefore be measured by \Gaia with a higher precision. This reflects in the fact that the distribution peaks to higher masses for lower relative error thresholds (brighter stars). In any case, we see that the shape of the curves resembles the ones obtained with the simple approach described in Section \ref{sec:mock} (see Fig. \ref{fig:MRmu}). 

We can compare our estimates with results from \cite{marchetti+17}, who data-mined \Gaia DR1/TGAS searching for HVSs. In the \textsc{Hills} catalogue we find a total of $5$ HVSs with a magnitude in the $V$ band lower than $11$, the $\sim 99\%$ completeness of the \emph{Tycho}-$2$ catalogue \citep{hog+00}. None of these stars are unbound, and the typical velocities are $< 400$ \kms.

\section{The \textsc{"MBHB"} Catalogue}  \label{sec:MBHB}

In this section, we explain how we create a mock population of HVSs ejected by a hypothetical massive black hole binary in the GC. We rely on results from full three-body scattering experiments presented in \cite{sesana+06}. In the following we will assume a massive black hole companion to Sagittarius A$^*$ with a mass $M_c = 5 \cdot10^3$ M$_\odot$, which can not be ruled out by the latest observational results of S stars in the Galactic Centre \citep{gillessen+17}. We assume a stellar density in the GC $\rho = 7 \cdot 10^4$ M$_\odot$ pc$^{-3}$ and a velocity dispersion of stars in the GC $\sigma = 100$ km s$^{-1}$ \citep{sesana+07}. The MBHB, with mass ratio $q \simeq 1.2 \cdot 10^{-3}$, is assumed to be in a circular orbit, with an initial separation $a_0 = 0.01$ pc at a given time $t_0$ after the Milky Was formed, corresponding to a look-back time $t_\mathrm{lb}$. Using the results presented in \cite{sesana+06}, we adopt the best-fit parameters for the lowest mass ratio explored in their simulation, i.e. $q = 1/243$. This choice is motivated by noticing that the authors' results do not vary appreciably when comparing results obtained for different mass ratios \cite[see Fig.3 and 5 in][]{sesana+06}. In the following we will assume that the orbit of the MBHB remains circular as the binary shrinks.

\subsection{Ejection of HVSs by the MBHB}

We create a grid of $100$ semi-major axes evenly spaced on a logarithmic scale, from a minimum value equal to $0.01$ $a_h$, to a maximum value of $a_0$. The value $a_h$ defines the minimum separation of a hard binary \citep{quinlan96}:
\begin{equation}
\label{eq:ah}
a_h = \frac{GM_c}{4 \sigma^2} \simeq 110 \ \mathrm{au} .
\end{equation}
The total stellar mass ejected by the binary in each bin is computed as \cite{sesana+06}:
\begin{equation}
\label{eq:deltaMej}
\Delta M_\mathrm{ej} = J (M_\bullet + M_c) \Delta \ln\Biggl(\frac{a_h}{a}\Biggr),
\end{equation}
where $a$ is the semi-major axis of the MBHB, and the mass ejection rate $J = J(a)$ is computed using the fitting function presented in \cite{sesana+06}, with best-fit parameters for a circular orbit with mass ratio $q = 1/243$.

\subsubsection{Rates of Orbital Decay}
\label{sec:rates_orb}

We now compare the rate of orbital decay of the MBHB due to the ejection of HVSs to the one due to the emission of gravitational waves (GWs). We determine the hardening rate of the binary following \cite{quinlan96}:
\begin{equation}
\label{eq:hardening}
H = \frac{\sigma}{G\rho}\frac{d}{dt}\Biggl(\frac{1}{a}\Biggr).
\end{equation}
A hard binary ($a <a_h$) hardens at a constant rate $H$. 

The rate of orbital decay due to the ejection of HVSs is then computed as:
\begin{equation}
\label{eq:dadt_HVS}
\frac{da}{dt}\Bigg|_\mathrm{HVS} = -\frac{G\rho H}{\sigma}a^2,
\end{equation}
where the hardening rate $H = H(a)$ is computed using the numerical fit in \cite{sesana+06} assuming a circular binary with $q = 1/243$.

The rate of orbital decay due to the emission of gravitational radiation can be approximated by \citep{peters64}:
\begin{equation}
\label{eq:dadt_GW}
\frac{da}{dt}\Bigg|_\mathrm{GW} =-\frac{64}{5}{G^3}{c^5}\frac{(M_\bullet M_c)(M_\bullet + M_c)}{a^3}.
\end{equation}
The two rates of orbital decay are equal for $\bar{a} = 48.4$ au $\sim 0.44 a_h$. For $a < \bar{a}$ the orbital evolution is dominated by the emission of gravitational waves, driving the binary to the merging. The binary will start evolve more rapidly, ejecting stars with a lower rate, since the time the binary spends in each bin of $a$ will be dictated by the emission of GWs. For $a < \bar{a}$ we therefore correct equation \eqref{eq:deltaMej} by multiplying it for $T_\mathrm{GW} / T_\mathrm{HVS}$, where $T_\mathrm{GW}$ is the time needed to shrink from $a$ to $a - \Delta a$ because of GWs emission, while $T_\mathrm{HVS}$ is the time the binary would have taken if it was driven by hardening. The times $T_\mathrm{HVS}$ and $T_\mathrm{GW}$ are computed, respectively, integrating equations \eqref{eq:dadt_HVS} and \eqref{eq:dadt_GW}.

\subsubsection{Creating the Mock Catalogue}

For each ejected mass bin $\Delta M_\mathrm{ej}$, see equation \eqref{eq:deltaMej}, we derive the corresponding number of HVSs $\Delta N$ as:
\begin{equation}
\label{eq:NHVS}
\Delta N = \frac{\Delta M_\mathrm{ej}}{\bigint_{M_\mathrm{min}}^{M_\mathrm{max}} M f(M) dM},
\end{equation}
where $f(M)$ is the stellar mass function in the GC, $M_\mathrm{min} = 0.1$ M$_\odot$, and $M_\mathrm{max} = 100$ M$_\odot$. We then draw $\Delta N$ stars of mass $M$ from a power-law mass function $f(M)$. 

\begin{figure}
	\centering
	\includegraphics[width=0.5\textwidth]{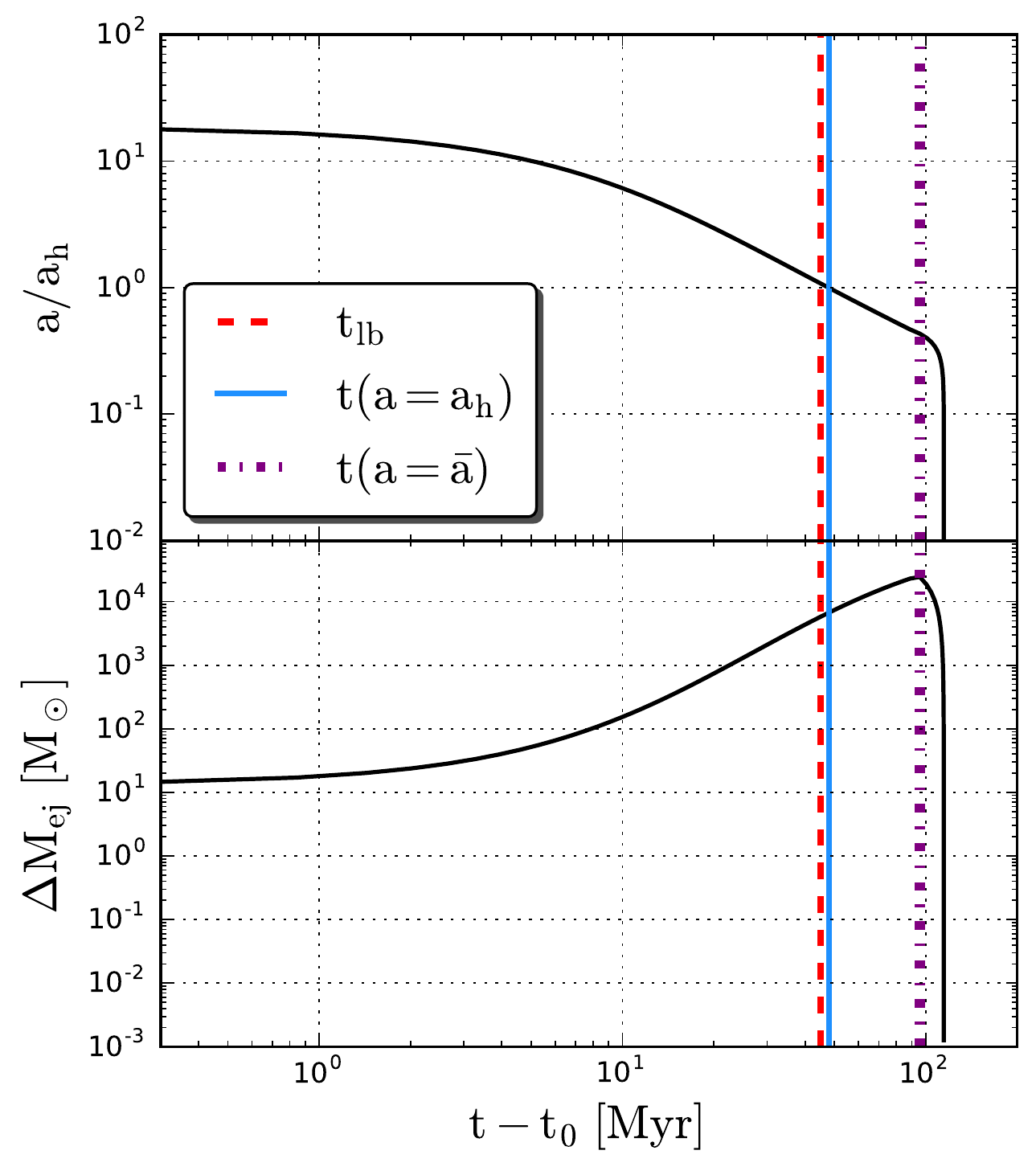}
	\caption{Time evolution of the MBHB binary separation (in units of $a_h$, top panel), computed integrating equations \eqref{eq:dadt_HVS} and \eqref{eq:dadt_GW}, and of the ejected stellar mass (bottom panel), computed using equation \eqref{eq:deltaMej}.}
	\label{fig:MBHB_aDeltaM}
\end{figure}

\begin{figure*}
	\centering
	\includegraphics[width=0.75\textwidth]{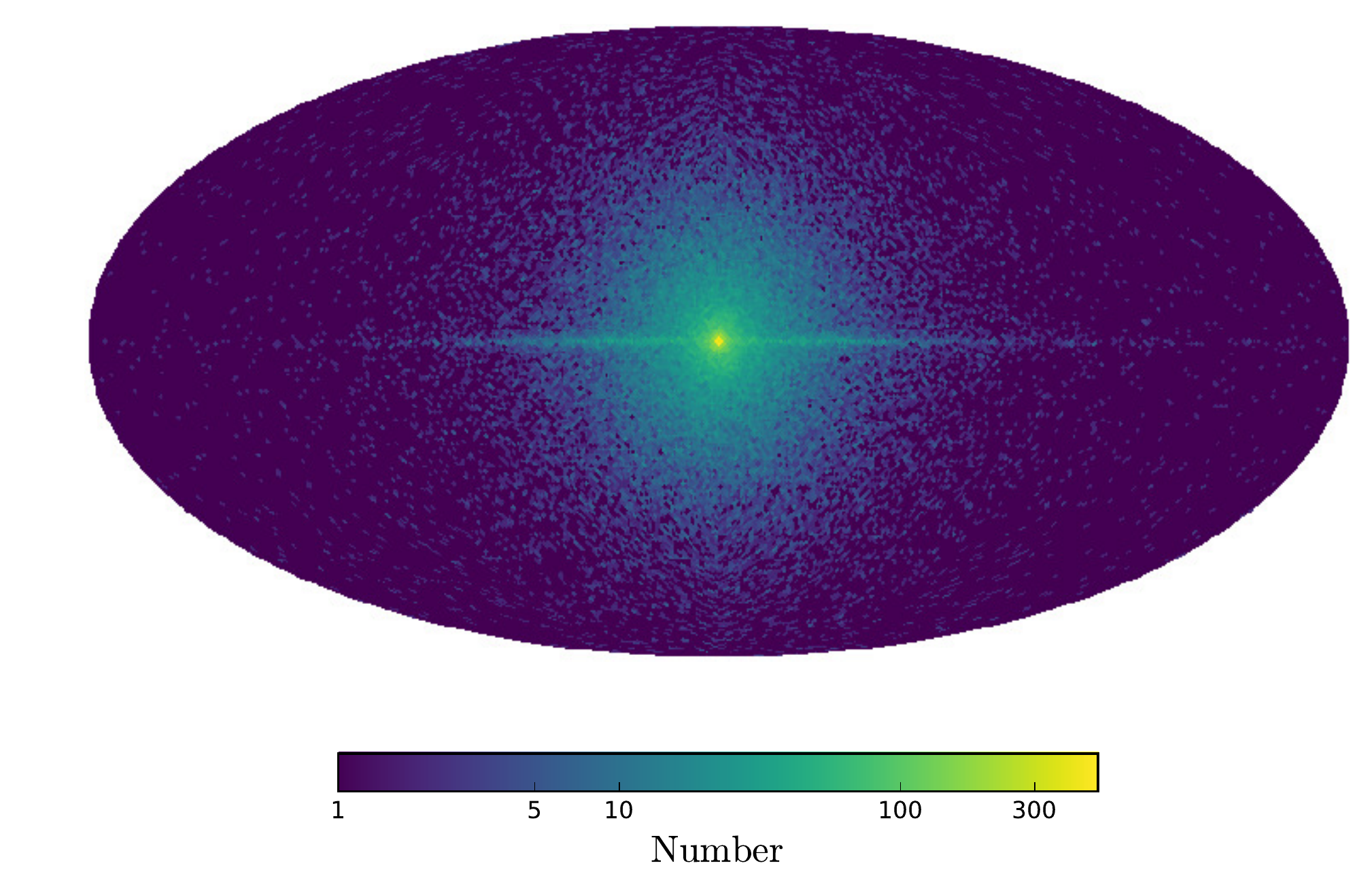}
	\caption{\textsc{MBHB} catalogue: sky distribution in Galactic coordinates of the current population of HVSs in our Galaxy ($122473$ stars).}
	\label{fig:MBHB_allHVSs}
\end{figure*}

\begin{figure*}
	\centering
	\includegraphics[width=0.8\textwidth]{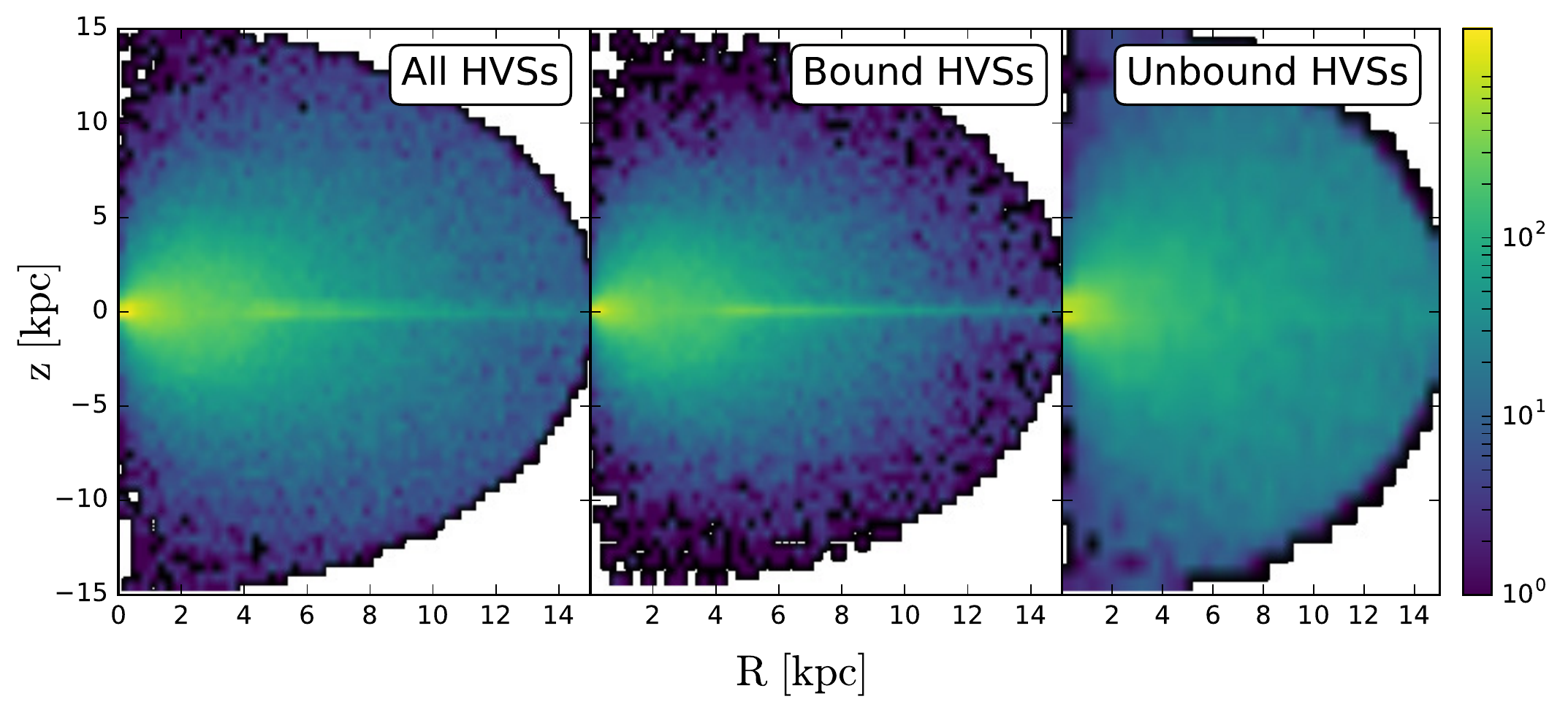}
	\caption{\textsc{MBHB} catalogue: distribution in Galactocentric cylindrical coordinates $(R, z)$ of all HVSs (left), bound HVSs (centre), and unbound HVSs (right) within $15$ kpc from the Galactic Centre.}
	\label{fig:MBHB_Rz}
\end{figure*}

We draw velocities from the velocity distribution \citep{sesana+06}:
\begin{equation}
\label{eq:MBHB_v}
f(w) = \frac{A}{h}\Biggl(\frac{w}{h}\Biggr)^\alpha \Biggl[1 + \Biggl(\frac{w}{h}\Biggr)^\beta\Biggr]^\gamma,
\end{equation}
where $w \equiv v/v_c$, $v_c = \sqrt{G(M_\bullet + M_c)/a}$ is the binary orbital velocity, $h \equiv \sqrt{2q}/(1+q)$, $A = 0.236$, $\alpha = -0.917$, $\beta = 16.365$, and $\gamma = -0.165$ \citep{sesana+06}. We note that in this scenario the ejection velocity does not depend on the mass of the HVS. We sample this velocity distribution using the MCMC sampler \textsc{emcee} \cite{emcee}. Velocities are drawn in the range $[v_\mathrm{min}, v_\mathrm{max}]$, $v_\mathrm{max} = v_c / (1 + q)$ \citep{sesana+06}. We fix $v_\mathrm{min}$ considering that we are only interested in stars with a velocity high enough to escape from the MW bulge. To be more quantitative, we only consider stars with a velocity $v$ greater then the escape velocity from the radius of influence of the binary, $r_\mathrm{inf} \equiv 2GM/(2\sigma^2) \sim 1$ pc. Assuming the same bulge profile as discussed in Section \ref{sec:mock:astro}, we get $v_\mathrm{min} = 645$ km s$^{-1}$, $\sim 100$ km s$^{-1}$ higher than the one used in \cite{sesana+06}. We note that since $a$ decreases with time, $v_c$ (and therefore $v_\mathrm{max}$) increase as the binary shrinks: HVSs with the highest velocities will be ejected right before the merger of the two black holes, but the majority of HVSs will be ejected right before the rate of orbital decay is driven by GW emission (see discussion in Section \ref{sec:rates_orb}).

For each star, we can compute the corresponding time of ejection after $t_0$: $\Delta t = t - t_0$, by integrating equation \eqref{eq:dadt_HVS} (equation \eqref{eq:dadt_GW}) for $a > \bar{a}$ ($a < \bar{a}$). The flight time of a star is computed according to $t^\prime = t_\mathrm{lb} - \Delta t$. The value of $t_\mathrm{lb}$ is chosen in such a way to match the observational estimate of $300$ HVSs in the mass range $[2.5, 4]$ M$_\odot$ within $100$ kpc from the GC. We find that we can match this value by assuming that the binary started to eject HVSs $t_\mathrm{lb} = 45$ Myr ago (see discussion in Section \ref{sec:MBHB_estimates}). 

We then determine the initial condition of the orbit and we propagate each star in the Galactic potential, with the same procedure outlined in Section \ref{sec:orbits}. In doing that, we assume for simplicity that the ejection of HVSs by the MBHB is isotropic. Photometry for each star is computed as in Section \ref{sec:mock:photo}, using equation \eqref{eq:tageH_MW} to determine the age of each star, and \Gaia errors on astrometry are estimated following Section \ref{sec:mock_errors}.

The evolution of the MBHB binary is summarized in Fig. \ref{fig:MBHB_aDeltaM}, where we plot the binary separation (top panel) and the ejected stellar mass (bottom panel) as a function of time. We highlight three key moments in the evolution of the system: the time at which it becomes a hard binary $t(a = a_h)$ (solid line), the time at which its evolution is driven by GW emission $t(a = \bar{a})$ (dot-dashed line), and the present time $t_\mathrm{lb}$ (dashed line). We can see that, to reproduce the estimates on the current population of HVSs, we are assuming that the MBHB in the GC has not yet shrunk to the hardening radius $a_h$, and that its evolution is still driven by dynamical hardening. Once GW emission dominates, the two black holes merge in a few Myr.

\subsection{\textsc{"MBHB"} Catalogue: Number Estimates of HVSs in \Gaia}
\label{sec:MBHB_estimates}

\begin{figure}
	\centering
	\includegraphics[width=0.5\textwidth]{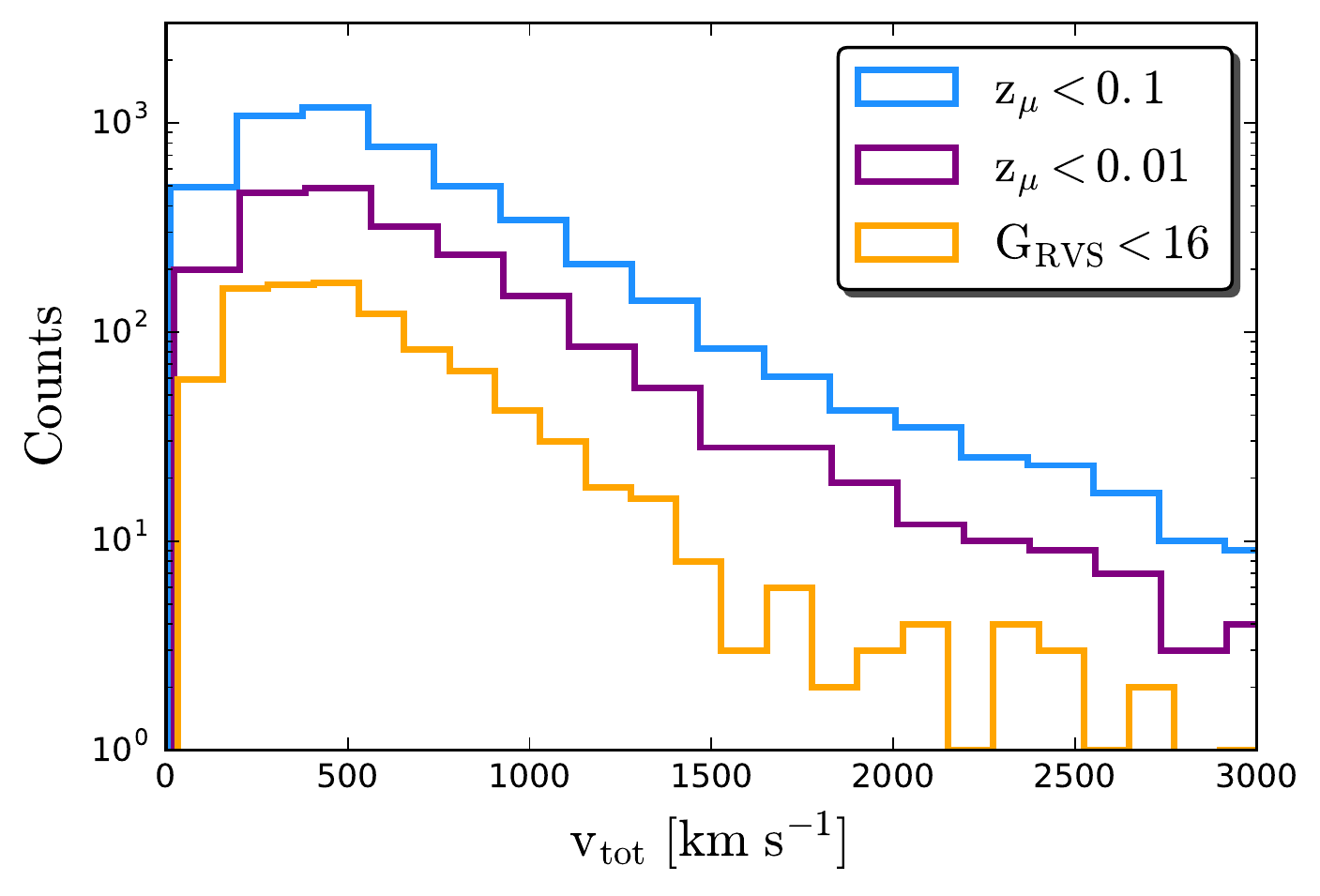}
	\caption{\textsc{MBHB} catalogue: total velocity (in the Galactocentric rest frame) of HVSs.}
	\label{fig:MBHB_vtotdistr}
\end{figure}

\begin{figure}
	\centering
	\includegraphics[width=0.5\textwidth]{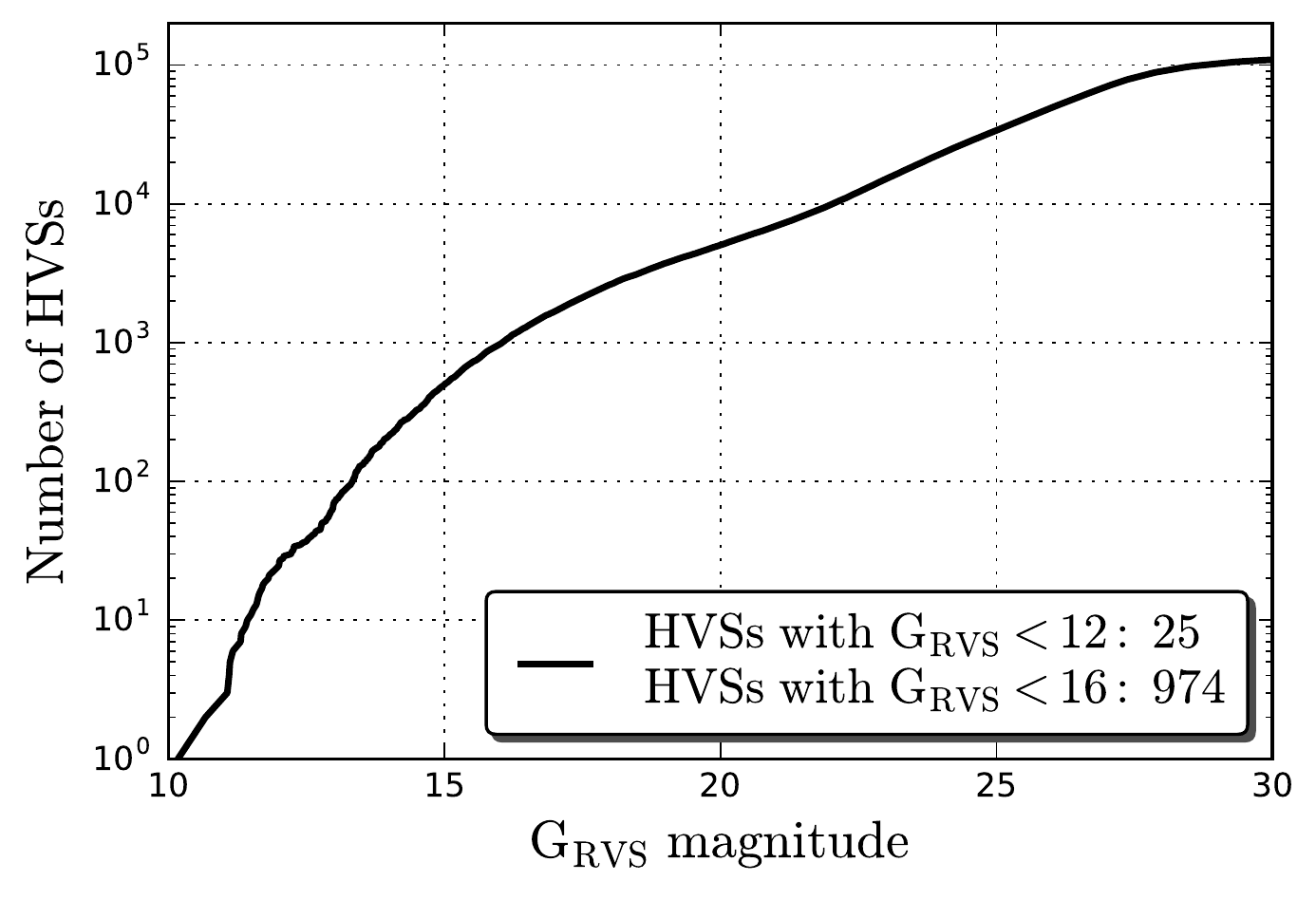}
	\caption{\textsc{MBHB} catalogue: cumulative distribution of HVSs magnitudes in the \Gaia $G_\mathrm{RVS}$ passband.}
	\label{fig:MBHB_Gdistr}
\end{figure}

Having created a catalogue of HVSs ejected by the MBHB, we can forecast how many of these HVSs we are expecting to find in the \Gaia catalogue. We find a total of $N = 122266$ HVSs ejected from the MBHB, corresponding to a total stellar mass $M_\mathrm{tot} \sim 3.7 \cdot 10^4$ M$_\odot$. We note that this number is about of the same order of magnitude than the estimate made using equation \eqref{eq:Ntot} for the \textsc{Hills} catalogue. 

The sky distribution of the population of HVSs is shown in Fig. \ref{fig:MBHB_allHVSs}. Fig. \ref{fig:MBHB_Rz} shows the distribution of stars within $15$ kpc from the GC in cylindrical coordinates $(R, z)$. We can see that the distribution of unbound HVSs is isotropic, while for bound HVSs the distribution is slightly tilted towards $z = 0$, because of the torque applied by the stellar disc.

We find $59$ \% of these stars to fly along bound orbits, and the total velocity distribution of the stars is shown in Fig. \ref{fig:MBHB_vtotdistr} for the subset of stars which will be precisely measured by \Gaia. Fig. \ref{fig:MBHB_Gdistr} shows the cumulative distribution of magnitudes in the \Gaia G$\mathrm{RVS}$ filter. A total of $974$ ($25$) stars will be brighter than than the $16$th ($12$th) magnitude, the magnitude limit for the final (second) data release of \Gaia. If we focus on the $G_\mathrm{RVS} < 16$ stars, we find that $328$ of them are unbound from the Milky Way, and that $527$ of them have a total velocity higher than $450$ \kms. We find $257$ unbound HVSs with mass between 2.5 and 4 MSun within 100 kpc from the GC, which agrees with the $300$ HVSs estimated in \cite{Brown+14} and the estimate presented in Section \ref{sec:Hillsestimates}. The distributions of errors in proper motions and parallax for the golden sample of HVSs with a three-dimensional velocity determination by \Gaia alone is shown in Fig. \ref{fig:err_bright}.

We predict $12$ of the $25$ $G_\mathrm{RVS} < 12$ stars to be unbound from the Galaxy. Their typical relative error in proper motions is $\lesssim 0.01\%$, and in parallax is $\lesssim 40\%$, with $80\%$ of the stars with $z_\varpi \lesssim 0.2$. These numbers have been corrected for the numerical factor introduced in Section \ref{sec:Hillsestimates}.

\begin{figure}
	\centering
	\includegraphics[width=0.5\textwidth]{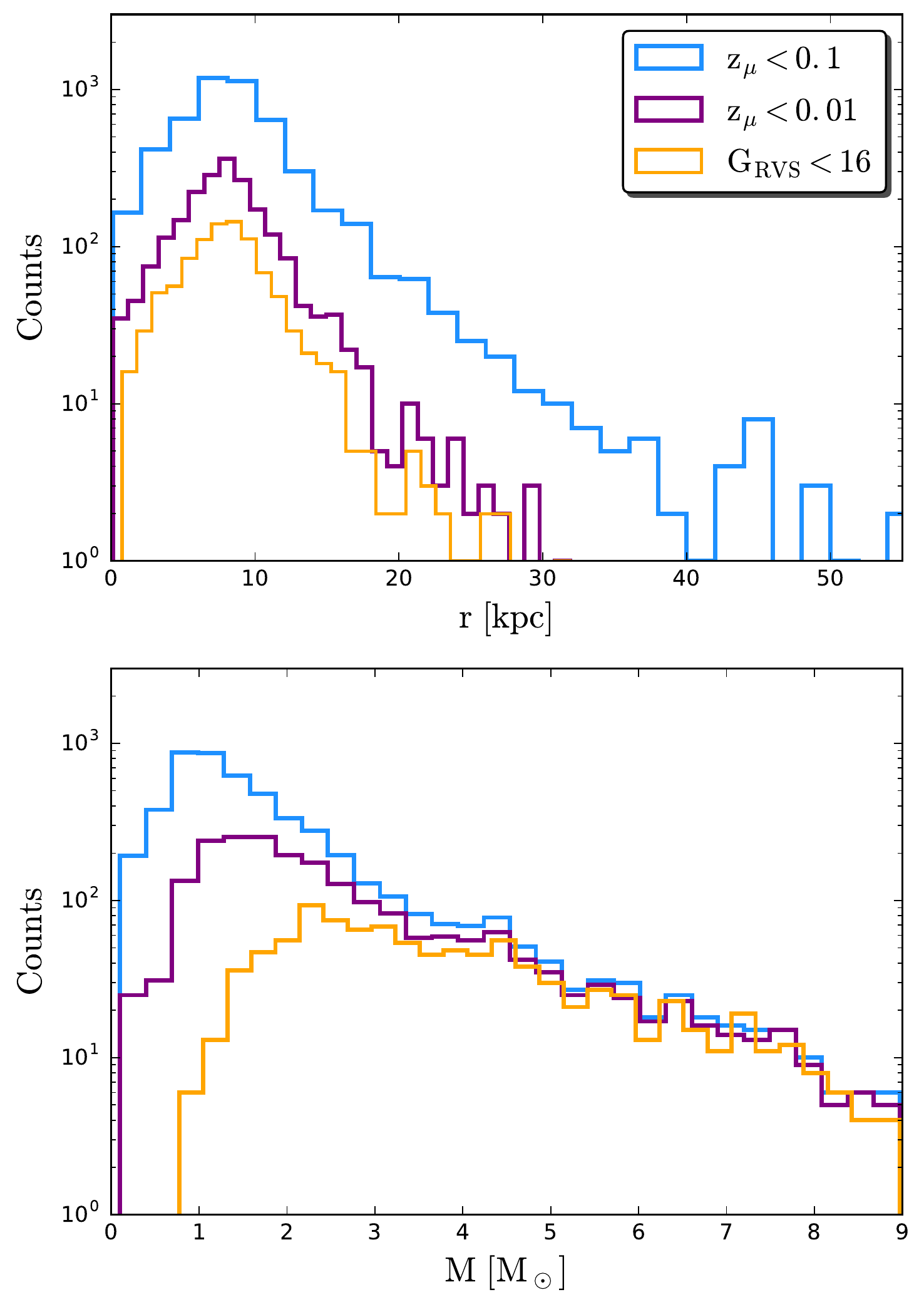}
	\caption{\textsc{MBHB} catalogue: heliocentric distance (top) and mass (bottom) distribution of the HVSs with a relative error on total proper motion below 10$\%$ (blue), 1$\%$ (purple), and with a radial velocity measurement by \Gaia (yellow).}
	\label{fig:MBHB_Mrdistr}
\end{figure}

The heliocentric distance (mass) distribution of HVSs in the catalogue with a precise astrometric determination by \Gaia is shown in the top (bottom) panel of Fig. \ref{fig:MBHB_Mrdistr}. Comparing these curves with the one obtained for the other mock catalogues, we can see that the shapes and the peak are reasonably similar, since they are shaped by the adopted mass function and stellar evolution model.

We can compare once more our estimates with results in \cite{marchetti+17} for \Gaia DR1/TGAS. We find a total of $2$ HVSs with $V < 11$. Both of these stars are unbound from the MW.

\section{Prospects for the Current Sample of HVSs} \label{sec:observations}

\begin{figure}
\centering
\includegraphics[width=0.5\textwidth]{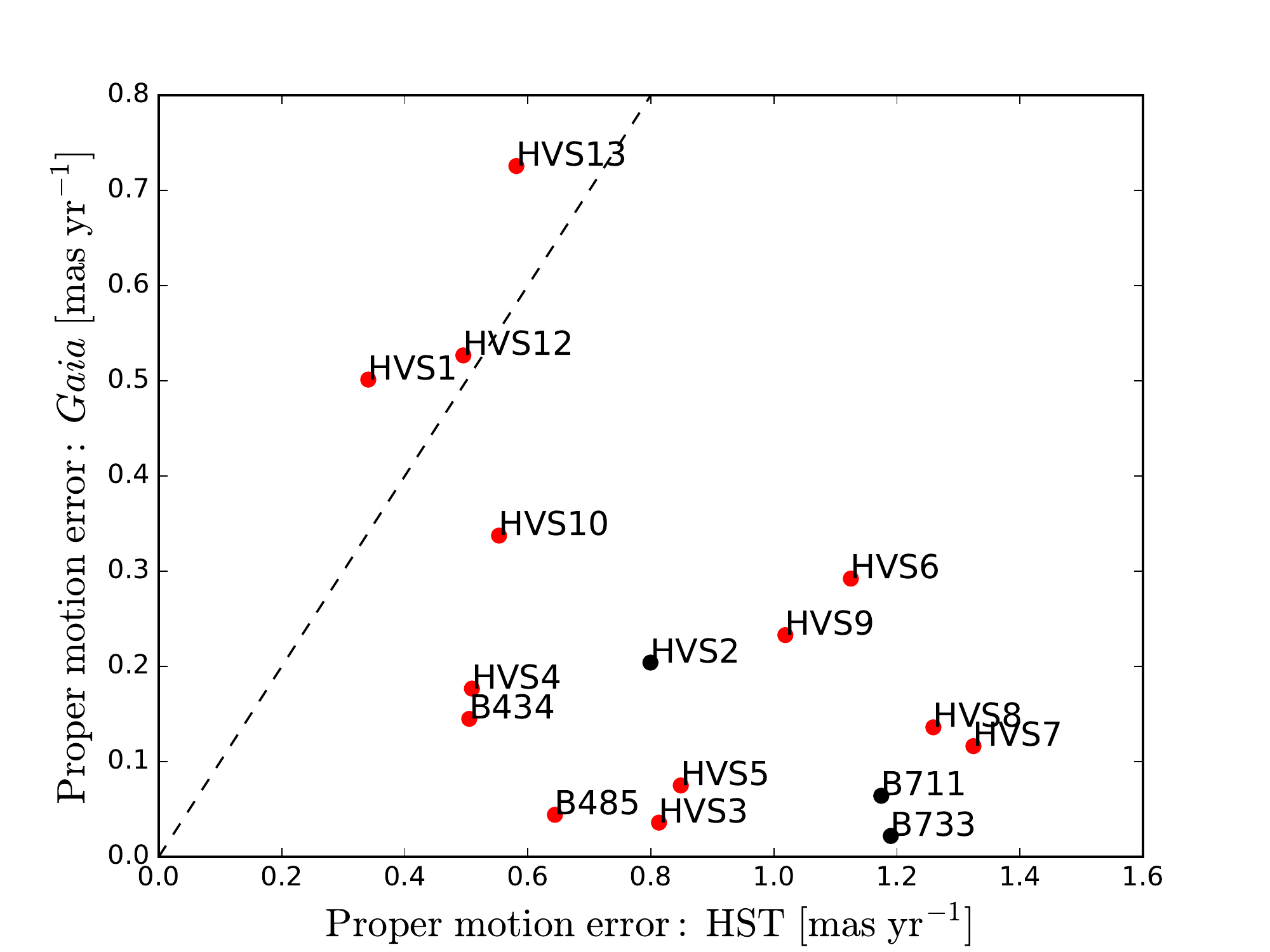}
\caption{Expected performance of \Gaia in measuring proper motions of the observed sample of candidates in \protect\cite{brown+15}. Red dots correspond to stars with a trajectory consistent with a GC origin, while black dots are disk runaways. On the $x$ axis we report the quadrature sum of the HST proper motion errors (Table 1 in \protect\cite{brown+15}), while on the $y$ axis the estimate obtained with \textsc{PyGaia}. Stars below the dashed line ($y=x$) will have a more precise proper motion determination in the final data release of the \Gaia mission.}
\label{fig:current_sample}
\end{figure}

In this section we assess the performance of \Gaia in measuring the astrometric properties of the current observed sample of HVS candidates. \cite{brown+15} measured proper motions with the \emph{Hubble Space Telescope} (HST) for $16$ extreme radial velocity candidates, finding that $13$ of them have trajectories consistent with a GC origin within $2\sigma$ confidence levels, and $12$ of them are unbound to the Milky Way. Proper motion accuracy is essential in constraining the origin of HVSs and is the main source of uncertainty in the orbital traceback, therefore we estimate \Gaia errors on the total proper motion for this sample of HVS candidates.

For each star we determine the ecliptic latitude using equation \eqref{eq:ecliptic}. We find $10$ of these $16$ stars in \Gaia DR1, from where we take \Gaia $G$ band magnitudes. All of the other stars but one \citep[HE $0437$-$5439$ = HVS$3$,][]{Edelmann+05} have SDSS magnitudes, and we compute \Gaia $G$ band magnitudes according to the polynomial fitting coefficients in \cite{jordi+10}. Conversion from SDSS passbands to $(V-I_c)$ Johnson-Cousins color index is done using the fitting formula in \cite{jordi+05}. For HVS$3$, we estimate the $G$ magnitude and the $(V-I_c)$ color from its $B$ and $V$ magnitude, according to \cite{natali+94, jordi+10}. We then use \textsc{PyGaia} to estimate \Gaia end-of-mission errors on the two proper motions for each star.

Fig. \ref{fig:current_sample} shows the comparison between HST proper motions determination and \Gaia estimates. In both cases we show the quadrature sum of the errors in the two proper motions. Stars with measurements consistent with coming from the GC are shown as red dots, while disk runaways are indicated as black dots, according to the classification presented in \cite{brown+15}. The black dashed line divides stars that will be detectable with a better accuracy than the current one: all stars but three (HVS$1$, HVS$12$, and HVS$13$) will have a better proper motion determination by \Gaia. This will help reducing in size the errorbars and identifying the ejection location, confirming or rejecting the GC origin hypothesis. This will be crucial to test the alternative ejection model presented in \cite{boubert+16, boubert+17}, where HVSs originate in the LMC.

We want once more to stress that these estimates refer to the final data release of the \Gaia satellite, currently planned for $2022$. Rescaling proper motion errors for the correcting factor $\sim 4.8$ introduced in Section \ref{sec:Hillsestimates}, we find that $7$ stars (the brightest in the sample) will have a better proper motion determination already in \Gaia DR2: HVS3, HVS5, HVS7, HVS8, B485, B711, and B733.

\section{Discussion and Conclusions} \label{sec:discussion}

In this paper we build mock catalogues of HVSs in order to predict their number in the following data releases of the \Gaia satellite. In particular, we simulate $3$ different catalogues:

\begin{enumerate}
	
	\item The \textsc{Vesc} catalogue does not rely on any assumption on the ejection mechanism for HVSs. We populate the Milky Way with stars on radial trajectories away from the Galactic Centre, and with a total velocity equal to the escape velocity from the Galaxy at their position. Therefore we only rely on the definition of HVSs as unbound stars, and we do not make any assumption on the physical process causing their acceleration. We then spatially distribute these stars assuming a continuous and isotropic ejection from the GC.
	
	\item The \textsc{Hills} catalogue focuses on the Hills mechanism, the leading mechanism for explaining the origin of HVSs. Assuming a parametrization of the ejection velocity distribution of stars from the GC, we numerically integrate each star's orbit, and we self consistently populate the Galaxy with HVSs.
	
	\item The \textsc{MBHB} catalogue assumes that HVSs are the result of the interaction of single stars with a massive black hole binary, constituted by Sagittarius A$^*$ and a companion black hole with a mass of $5 \cdot 10^3$ M$_\odot$. In this and in the previous catalogue there are bound HVSs: stars that escape the GC with a velocity which is not high enough to escape from the whole Galaxy. These are the result of modelling a broad ejection velocity distribution.
	
\end{enumerate}

We characterize each star in each catalogue from both the astrometric and photometric point of view. We then derive the star magnitude in the \Gaia passband filters and the \Gaia measurement errors in its astrometric parameters. The aim is to assess the size and quality of the \Gaia HVS sample.

As a summary and for quick consultation, our results for the size of three mock catalogues discussed in the paper are summarised in Table \ref{TAB:summary} for the final \Gaia data release, and in Table \ref{TAB:summary_DR2} for the second data release. Regardless of the adopted ejection mechanism, we can conclude that \Gaia will provide an unprecedented sample of HVSs, with numbers ranging from several hundreds to several thousands. The peak of the mass distribution and the limiting heliocentric distance at which HVSs will be observed by \Gaia are presented in Table \ref{TAB:summaryMR}. We can see that these values differ from the current sample of observed late B-type stars in the outer halo (refer also to Fig. \ref{fig:MRmu}, \ref{fig:Mrdistr}, \ref{fig:MBHB_Mrdistr}). Most HVSs will have precise proper motion measurements, and therefore data mining techniques not involving the radial velocity information need to be developed in order to extract them from the dominant background of other stars in the MW \citep{marchetti+17}. Stars with precise proper motions will be visible at typical heliocentric distances $r < 50$ kpc, while stars bright enough to have a radial velocity measurement from \Gaia will typically be observed at $r<30$ kpc, with a peak in the distribution for $r \sim 10$ kpc. 

We estimate the precision with which \Gaia will measure proper motions for the  sample of HVSs candidates presented in \cite{brown+15}. Fig. \ref{fig:current_sample} shows that the majority of HVSs will have a better proper motion determination by \Gaia. This will help determining their ejection location, confirming or rejecting the Galactocentric origin hypothesis. 

\begin{table*}
\caption{Number estimates of HVSs in the \emph{final} data release of \Gaia, for the three implemented catalogues of HVSs: \textsc{Vesc}, \textsc{Hills}, and \textsc{MBHB}. N$_\mathrm{tot}$ is the total number of HVSs in the Galaxy, N$(z_\mu < 0.1)$ (N$(z_\mu < 0.01)$) is the number of HVSs which will be detected by \Gaia with a relative error on total proper motion below 10\% (1\%), N$(z_\varpi < 0.2)$ is the number of HVSs with a relative error on parallax below 20\%, and N$_\mathrm{vrad}$ is the number of stars bright enough to have a radial velocity measurement. We remind the reader that the \textsc{Vesc} catalogue, by construction, only includes unbound objects, while the \textsc{Hills} and the \textsc{MBHB} catalogues contain both bound and unbound stars.}
\label{TAB:summary}

\begin{threeparttable}

\begin{tabular}{lccccc}
	
\hline

Catalogue & N$_\mathrm{tot}$ & N$(z_\mu < 0.1)$ & N$(z_\mu < 0.01)$ & N$(z_\varpi < 0.2)$ & N$_\mathrm{vrad}$  \\

\hline

\textsc{Vesc} & $17074$ & $709$ & $241$ & $40$ & $115$  \\

\textsc{Hills} & $100000$ & $11661$ & $3765$ & $568$ & $2140$ \\

\textsc{MBHB} & $122266$ & $5066$ & $2124$ & $364$ & $974$  \\

\hline

\end{tabular}

\end{threeparttable}

\end{table*}

\begin{table*}
\caption{Same as \ref{TAB:summary}, but for predictions of HVSs in the \emph{second} data release of \Gaia.}
\label{TAB:summary_DR2}

\begin{threeparttable}

\begin{tabular}{lccccc}
	
\hline

Catalogue & N$_\mathrm{tot}$ & N$(z_\mu < 0.1)$ & N$(z_\mu < 0.01)$ & N$(z_\varpi < 0.2)$ & N$_\mathrm{vrad}$ \\

\hline

\textsc{Vesc} & $17074$ & $357$ & $81$ & $20$ & $2$ \\

\textsc{Hills} & $100000$ & $5963$ & $781$ & $261$ & $19$  \\

\textsc{MBHB} & $122266$ & $2892$ & $750$ & $194$ & $25$ \\

\hline

\end{tabular}

\end{threeparttable}

\end{table*}

\begin{table*}
\caption{Peak mass of the mass distribution and maximum heliocentric distance for the HVSs in the three different mock catalogues. The maximum heliocentric distance is defines as the distance at which we predict a total of $0.5$ stars. Due to the small number of HVSs with a three-dimensional velocity in \Gaia DR2, we choose not to characterize their distributions here.}
\label{TAB:summaryMR}

\begin{threeparttable}

\begin{tabular}{lcccc}
	
\hline

Catalogue & $z_\mu < 0.1$ & $z_\mu < 0.01$ & $z_\varpi < 0.2$ & vrad  \\

\hline

\textsc{Vesc} &  ($1.0$ M$_\odot$, $40$ kpc) & ($1.5$ M$_\odot$, $25$ kpc) & ($2.5$ M$_\odot$, $12$ kpc) & ($2.7$ M$_\odot$, $25$ kpc)  \\

\textsc{Hills} & ($1.2$ M$_\odot$, $48$ kpc) & ($2.1$ M$_\odot$, $20$ kpc) & ($2.9$ M$_\odot$, $10$ kpc) & ($3.0$ M$_\odot$, $18$ kpc)  \\

\textsc{MBHB} & ($0.8$ M$_\odot$, $41$ kpc) & ($1.4$ M$_\odot$, $28$ kpc) & ($1.5$ M$_\odot$, $12$ kpc) & ($2.3$ M$_\odot$, $24$ kpc)  \\

\hline

\end{tabular}

\end{threeparttable}

\end{table*}

We now briefly discuss the impact of the assumptions made on the stellar population in the GC. The \textsc{Vesc} catalogue does not depend on the binary population properties, but only on the choice of the Galactic potential, which we fix to a fiducial model consistent with the latest observational data on the rotation curve of the MW. In the \textsc{Hills} catalogue, our choice for the binary distribution parameters $\alpha=-1$, $\gamma=-3.5$ is motivated by the fit of the sample of unbound late B-type HVSs to the velocity distribution curve modelled using the Hills mechanism \citep{Rossi+17}. We repeat the same analysis presented in Section \ref{sec:Hills} adopting $\gamma=0$: a flat distribution of binary mass ratios. This choice implies a higher mass for the secondary star in the binary, compared to the steeper value of $\gamma = -3.5$. Given the mass dependency of equation \eqref{eq:vdistr}, this results in high total velocities for binaries in which the HVSs is the primary star. This in turn implies, on average, a larger number of HVSs with higher mass, which will be observed by \Gaia to higher heliocentric distances with lower relative errors. Nevertheless, the final estimates on the number of HVSs we are expecting to be found in the \Gaia catalogue are consistent with results presented in Section \ref{sec:Hills}. A choice of a top-heavy initial mass function for stars in the GC \citep[e.g.][]{bartko+10, lu+13} would produce similar results. 
As a further check, we study the impact of adopting Galactic binary properties, which can be significantly different than in star forming regions, such as $30$ Doradus in the LMC or the GC \citep{D&K13, sana+13, kobulnicky+14}. In particular, we choose to change our prescription for solar mass HVSs, which are the majority of stars in our simulations. From equation \eqref{eq:vdistr}, we can see that, for an equal mass binary ($q=1$) with $M = 1$ M$_\odot$, the maximum initial separation needed in order to attain ejection velocity of $680$ \kms is $a_\mathrm{max} \sim 100$ R$_\odot$. This choice of ejection velocity, given our adopted model for the Galactic potential, is the minimum velocity needed for a star in the GC to reach the Sun position with zero velocity. This maximum binary separation corresponds to a maximum orbital period $P_\mathrm{max} \sim 90$ days. For solar-type primaries ($m_p < 1.2$ M$_\odot$) in binaries with periods shorter than $P_\mathrm{max}$, the mass ratio distribution can be approximated as a broken power-law, with indexes $\gamma_{\mathrm{small}q} = 0.3$ (for $0.1 < q < 0.3$) and $\gamma_{\mathrm{large}q}\ = -0.5$ (for $0.3 < q < 1.0$) \citep{moe+17}. The period distribution is flat with very good approximation in this restricted period range \cite[see Figure 37 in][]{moe+17}. Moreover, solar mass stars are single twice as often as $B$-type stars \citep{moe+17}, therefore, when we draw primary masses from the Kroupa mass function, we select stars with $m_p < 1.2$ M$_\odot$ only $50 \%$ of the times. With these prescriptions, using equation \eqref{eq:Ntot} with this updated $dN/dM$ we again obtain $N_\mathrm{tot} \simeq 1\cdot 10^5$. Because of the lower number of solar mass stars in binary systems, we now find the mass distribution to peak around $1.5$ M$_\odot$ for stars with precise proper motions by \Gaia. Apart from this, number estimates agree extremely well with results presented in Section \ref{sec:Hillsestimates}.
Constructing the \textsc{MBHB} catalogue it is also worth exploring different values for the mass of the secondary black hole, which we fixed to $5\cdot 10^3$ M$_\odot$. Higher (lower) masses result in a larger (smaller) total mass ejected by the binary (see equation \eqref{eq:deltaMej}). Tuning the value of $t_\mathrm{lb}$, the loockback time at which the MBHB started ejecting HVSs, it is then possible to find different values of the secondary mass which are consistent with the observational estimate given by \cite{Brown+14}. Regardless of $t_\mathrm{lb}$, we find $M_c = 1000$ M$_\odot$ to be a lower limit on the black hole mass to be able to observe $300$ HVSs in the observed mass range $[2.4, 5]$ M$_\odot$, within $300$ kpc from the GC. The possibility of considering multiple merging events, and/or a full parameter space exploration to break the degeneracy between $M_c$ and $t_\mathrm{lb}$ are beyond the scope of this paper. An improvement over this catalogue would consist in modelling the ejection angles of HVSs as a function of the decreasing binary separation. 

Although a full investigation of the detection strategy of HVSs is beyond the scope of this paper, it is interesting to qualitatively compare our findings with the expected major sources of sample contamination. HVSs may be confused with \emph{runaway} stars: stars ejected with high velocities by dynamical encounters in dense stellar systems \citep{poveda+67, portegieszwart00} or by the explosion of a supernova in a binary star \citep{blaauw61, tauris+98}. These stars are produced in star forming regions in the stellar disk of the Milky Way, but, given their high velocity, they can travel to the stellar halo \citep{silva+11}. The \Gaia catalogue will contain $\sim 10^9$ disk stars \citep{robin+12}. Assuming rates and the velocity distribution in \cite{silva+11}, we can estimate a total of $N_\mathrm{RS} \sim 10^5$ runaway stars in the \Gaia catalogue with $v > 400$ \kms, two order of magnitudes more than the predicted number of HVSs. Nevertheless, the rate of ejection of unbound objects is estimated to be approximately one for every $100$ HVSs \citep{brown15}, with velocities that can reach up to $\sim 1300$ \kms for companion stars in a binary disrupted via an asymmetric supernova explosion \citep{tauris15}. Precise proper motions and radial velocities provided by \Gaia will help discriminating these stars, by tracing back their orbits to determine the ejection location (GC or stellar disk). High velocity halo stars on radial orbits could also be easily mistaken for bound HVSs because of their similar trajectories. To estimate the contamination of such stars to the sample of bound HVSs, we start considering that we are expecting $\sim 10^7$ halo stars in the \Gaia catalogue \citep{robin+12}. We estimate a total of $\sim 10^5$ halo stars with a total velocity vector pointing inside the solid angle subtending a cone with base radius of $500$ pc around the GC when traced back in time. Given the typical velocity dispersion of stars in the stellar halo $\sim\sqrt{3} \cdot 150$ \kms \citep{smith+09, evans+16}, we expect $\sim 2000$ halo stars on radial trajectories from the GC with $v > 400$ \kms. Further stellar properties, such as metallicity, need to be considered in order to correctly classify those stars \citep[e.g.][]{hawkins+15, zhang+16}.

To summarize, the sample of known HVSs will start increasing in number in April 2018 with DR2, with a few tens of stars with a precise three-dimensional velocity by \Gaia alone. This sample will already be comparable in size with the current tens of HVSs candidates, but the largest improvement in terms of stars with full three-dimensional velocity will come with the final \Gaia data release, with hundreds of stars unbound from the Milky Way. The majority of HVSs in \Gaia will not have radial velocities from {\it Gaia}, therefore dedicated spectroscopic follow-up programs with facilities such as 4MOST \citep{dejong+16} and WEAVE \citep{dalton16} will be necessary to derive their total velocity and to clearly identify them as HVSs.

\section*{Acknowledgements}

We thank Warren Brown for the careful reading of the manuscript and the useful comments. TM and EMR acknowledge support from NWO TOP grant Module 2, project number 614.001.401. AS is supported by the Royal Society. This research made use of \textsc{Astropy}, a community-developed core \textsc{Python} package for Astronomy \citep{astropy}. All figures in the paper were produced using \textsc{matplotlib} \citep{matplotlib}. This work would not have been possible without the countless hours put in by members of the open-source community all around the world.

\bibliographystyle{mnras}
\bibliography{Gaia_predictions}

\bsp

\label{lastpage}

\end{document}